\documentclass[journal]{IEEEtran}
\usepackage[utf8]{inputenc}

\usepackage{color}
\usepackage[pdftex]{graphicx}
\usepackage{balance}
\usepackage{caption}
\usepackage{subcaption}
\usepackage{textcomp}
\usepackage{float}
\usepackage{tabularx}
\usepackage{multirow}
\usepackage{amsthm}
\usepackage[table]{xcolor}
\usepackage[acronym,nopostdot]{glossaries}
\usepackage[noabbrev]{cleveref}

\definecolor{babyblueeyes}{rgb}{0.6, 0.7, 0.9}

\definecolor{aliceblue}{rgb}{0.8, 0.9, 1.0}

\title{Location Management in IP-based Future LEO Satellite Networks: A Review }

\author{Tasneem Darwish,                               ~\IEEEmembership{Senior Member,~IEEE,}         Gunes Karabulut Kurt, 
        ~\IEEEmembership{Senior Member,~IEEE,} Halim Yanikomeroglu, ~\IEEEmembership{Fellow, ~IEEE,} Guillaume Lamontagne, and Michel Bellemare
\thanks{Tasneem Darwish and Halim Yanikomeroglu are with the Department of Systems and Computer Engineering, Carleton University, Ottawa, Canada (e-mail: tasneemdarwish@sce.carleton.ca; halim@sce.carleton.ca).\protect\\
Gunes Karabulut Kurt is with the Poly-Grames Research Center, Department of Electrical Engineering,  Polytechnique Montr\'eal, Montr\'eal, Canada (e-mail: gunes.kurt@polymtl.ca). \protect\\
Guillaume Lamontagne and Michel Bellemare are with the Division of Satellite Systems, MDA, Canada (e-mail: Guillaume.Lamontagne@mda.space; Michel.Bellemare@mda.space).

}
}

\date{January 2021}

\makeglossaries 
\newacronym{ioe}{IoE}{Internet of Everything}
\newacronym{iot}{IoT}{Internet of Things}
\newacronym{3gpp}{3GPP}{3rd Generation Partnership Project}
\newacronym{satnet}{SatNet}{Satellite Network}
\newacronym{leo}{LEO}{Low Earth Orbit}
\newacronym{obp}{OBP}{On-board Processor}
\newacronym{meo}{MEO}{Medium Earth Orbit}
\newacronym{geo}{GEO}{Geostationary Orbit}
\newacronym{ietf}{IETF}{Internet Engineering Task Force}
\newacronym{mip}{MIP}{Mobile Internet Protocol}
\newacronym{mipv4}{MIPv4}{Mobile Internet Protocol version 4}
\newacronym{mipv6}{MIPv6}{Mobile Internet Protocol version 6}
\newacronym{pmipv6}{PMIPv6}{Proxy Mobile Internet Protocol version 6}
\newacronym{tcp}{TCP}{Transport Control Protocol}
\newacronym{mn}{MN}{Mobile Node}
\newacronym{ap}{AP}{Access Point}
\newacronym{bs}{BS}{Base Station}
\newacronym{hip}{HIP}{Host Identity Protocol}
\newacronym{fmipv6}{FMIPv6}{Fast Handovers for Mobile Internet Protocol version 6}
\newacronym{hmipv6}{HMIPv6}{Hierarchical Mobile Internet Protocol version 6}
\newacronym{ha}{HA}{Home Agent}
\newacronym{coa}{CoA}{Care-of-Address}
\newacronym{bu}{BU}{Binding Update}
\newacronym{ba}{BA}{Binding Acknowledgement}
\newacronym{cn}{CN}{Corresponding Node}
\newacronym{ar}{AR}{Access Router}
\newacronym{far}{FAR}{Foreign Access Router}
\newacronym{par}{PAR}{Previous Access Router}
\newacronym{rs}{RS}{Router Solicitation}
\newacronym{pradv}{PRAdv}{Proxy Router Advertisement}
\newacronym{fbu}{FBU}{Fast Binding Update}
\newacronym{fna}{FNA}{Fast Neighbour Advertisement}
\newacronym{map}{MAP}{Mobility Anchor Point}
\newacronym{rcoa}{RCoA}{Regional Care-of-Address}
\newacronym{lcoa}{LCoA}{on Link Care-of-Address}
\newacronym{dad}{DAD}{Duplicate Address Detection}
\newacronym{irtf}{IRTF}{Internet Research Task Force}
\newacronym{lisp}{LISP}{Locator/Identity Separation Protocol}
\newacronym{eid}{EID}{End Identifier}
\newacronym{rloc}{RLOC}{Routing Locator}
\newacronym{itr}{ITR}{Ingress Tunnel Router}
\newacronym{etr}{ETR}{Egress Tunnel Router}
\newacronym{mag}{MAG}{Mobile Access Gateway}
\newacronym{lma}{LMA}{Local Mobility Anchor}
\newacronym{pbu}{PBU}{Proxy Binding Update}
\newacronym{pba}{PBA}{Proxy Binding Acknowledgement}
\newacronym{bce}{BCE}{Binding Cache Entry}
\newacronym{isl}{ISLs}{Inter Satellite Links}
\newacronym{ilnp}{ILNP}{Identifier Locator Network Protocol}
\newacronym{sdn}{SDN}{Software Defined Network}
\newacronym{dmm}{DMM}{Distributed Mobility Management}
\newacronym{lm}{LM}{Location Manager}
\newacronym{la}{LA}{Location Area}
\newacronym{fes}{FES}{Fixed Earth Station}
\newacronym{gs}{GS}{Ground Station}
\newacronym{gsl}{GSL}{Ground to Satellite Link}
\newacronym{vac}{VAC}{Virtual Agent Cluster}
\newacronym{vad}{VAD}{Virtual Agent Domain}
\newacronym{maa}{MAA}{Mobile Agent Anchor}
\newacronym{hmaa}{HMAA}{Home Mobile Agent Anchor}
\newacronym{tgms}{TGMS}{Terrestrial Gateway Mapping Server}
\newacronym{id}{ID}{Identifier}
\newacronym{vap}{VAP}{Virtual Attachment Point}
\newacronym{rmrs}{RMRS}{Rapid Mapping Resolution System}
\newacronym{haps}{HAPS}{High Altitude Platform System}
\newacronym{vhetnet}{VHetNet}{Vertical Heterogeneous Network}
\newacronym{td}{TD}{Topology Discovery}
\newacronym{icn}{ICN}{Information Centric Network}
\newacronym{sdsn}{SDSN}{Software Defined Satellite Network}
\newacronym{nocc}{NOCC}{Network Operation and Control Centre}
\newacronym{tcam}{TCAM}{Ternary Content Addressable Memory}
\newacronym{tsmm}{TSMM}{Timeout Strategy-based Mobility Management}
\newacronym{dct}{DCT}{Dynamic Classified Timeout}
\newacronym{ilp}{ILP}{Integer Linear Programming}
\newacronym{dcpp}{DCPP}{Dynamic Controller Placement Problem}
\newacronym{scpp}{SCPP}{Static Controller Placement Problem}
\newacronym{apso}{APSO}{Accelerated Particle Swarm Optimization}
\newacronym{qos}{QoS}{Quality of Service}

\begin{document}

\maketitle

\begin{abstract}
Future integrated terrestrial, aerial, and space networks will involve thousands of \acrfull{leo} satellites forming a network of mega-constellations, which will play a significant role in  providing communication and Internet services everywhere, at any time, and for everything. Due to its very large scale and highly dynamic nature, future LEO satellite networks (SatNets) management is a very complicated and crucial process, especially the mobility management aspect and its two components  location management and  handover management. In this article, we present a comprehensive and critical review of the state-of-the-art research in LEO \acrshort{satnet}s location management. First, we give an overview of the \acrfull{ietf} mobility management standards (e.g., Mobile IPv6 and Proxy Mobile IPv6) and discuss their location management techniques limitations in the environment of future LEO \acrshort{satnet}s. We highlight future LEO \acrshort{satnet}s mobility characteristics and their challenging features and describe two unprecedented future location management scenarios. A taxonomy of the available location management solutions for LEO \acrshort{satnet}s is presented, where the solutions are classified into three approaches. \textcolor{black}{The \textit{``Issues to consider"} section draws attention to critical points related to each of the reviewed approaches that should be considered in future LEO SatNets location management. To identify the gaps, the current state of LEO SatNets location management is summarized. Noteworthy future research directions are recommended.} This article is providing a road map for researchers and industry to shape the future of LEO SatNets location management. 
\end{abstract}


\glssetwidest{HAPS-SMBS}
\hspace{1cm}\printglossary[style=alttree,type=\acronymtype,title=Abbreviations,nogroupskip, nonumberlist]

\section{Introduction}
With the emergence of the \acrfull{ioe}  paradigm, which involves people, data, intelligent processes, sensors, and devices \cite{Miraz2015}, wireless communication networks are going through an unprecedented revolution to meet the requirements of \acrshort{ioe} global deployment. It is anticipated that future networks will have to ensure the provision of  communications and computation services, and security for a tremendous number of devices with very broad and demanding requirements in a ubiquitous manner. This fuels the need for providing broadband Internet connectivity everywhere on Earth and even within its surrounding space. Although terrestrial communication networks have witnessed several significant advances, the coverage of communication networks is still patchy, particularly in rural and difficult-to-serve areas (e.g., seas, oceans, polar regions, and high altitudes in the sky). During the COVID-19 pandemic, many people/companies realized that work can be done remotely and it is not necessary to go to working places. This might encourage many people to leave the big cities and move to live in more relaxing and less expensive areas (e.g., rural areas), even after the pandemic is over. In this situation, there will be new population distribution which requires the provision of the Internet in more scattered spots. Besides providing coverage to rural and difficult-to-serve areas, the large footprint of satellite networks can boost the communication capacity for a huge number of terrestrial users on a flexible basis. This makes satellites ideal for providing broadcasting or multicasting services.  In addition, satellite networks can offer critical and emergency services during and after natural disasters.  

Recently several industrial groups and standardization organizations, including the \acrfull{3gpp}, are proposing the integration of satellite networks with 5G and beyond to support seamless and broadband coverage everywhere, for everything, at any time \cite{Li2019}.
 Driven by the growing demands for Internet and communication, satellite networks have developed fast during the last 10 years \cite{Han2016b}, especially for the low Earth orbit (LEO) satellite networks (SatNets) such as the Iridium NEXT system and the upcoming SpaceX mega-constellations. The objective is to cover the entire Earth with LEO satellites equipped with \acrfull{obp} devices. Such SatNets can be considered as an extension of the terrestrial IP network or as a standalone satellite network where satellites are the data sources, processors, and consumers \cite{Shahriar2008}.


 Due to their low altitudes (160-2000 km), 
 LEO satellites provide low-latency communications in comparison to \acrfull{meo} and 
\acrfull{geo} satellites. However, the fast movement on the low Earth orbits comes with the price of the very frequent handovers/disconnections in communications with ground stations and users, aerial network entities, and other LEO satellites \cite{Shahriar2008}. 
For example, a LEO satellite at 500 km altitude travels at 7.6 km/s and it takes around 95 minutes to orbit the Earth resulting in a handover every 5 minutes approximately.
Therefore, there is a pressing need for efficient mobility management protocols to provide seamless communication between the satellite networks and the Internet \cite{Jiang2016}.

\begin{figure}[h]
\centering
\includegraphics[width=0.45\textwidth]{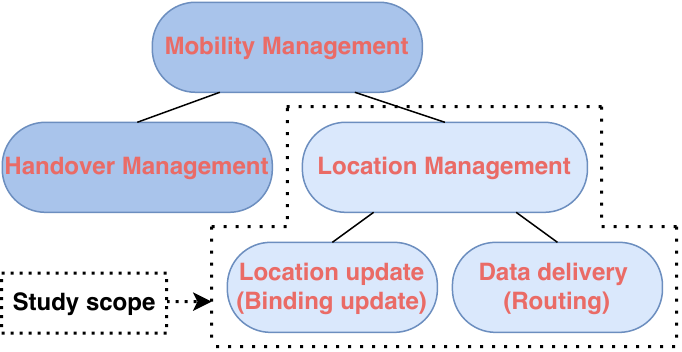}
\caption{The scope of this study.}
\label{Scope}
\end{figure}

\acrshort{ietf} introduced the \acrfull{mip} protocol (i.e., \acrshort{mipv4} and \acrshort{mipv6}) and \acrfull{pmipv6} protocol 
to provide mobility management in the terrestrial IP networks. IETF's IP-based mobility management protocols aim at maintaining the \acrfull{tcp} connection between a \acrfull{mn} and a static \acrfull{ap} or \acrfull{bs} while reducing the effects of location changes of the MN. This is achieved through the mobility management interrelated components, handover management and location management. Handover management is the process by which a MN keeps its connection active while moving from one AP to another. Location management has two components, location update which is the process of identifying and updating the logical location of the MN in the network, and the second component is data delivery (i.e., routing) which forwards the data packets directed to the MN to its new location. This study focuses on the location management side with its two components, location updates (binding updates) and data delivery (routing), as described in Figure \ref{Scope}.  


Due to the differences between terrestrial networks and SatNets in terms of topology, processing power, and communication links, the application of standard IP mobility management protocols, and more specifically their location management techniques, to satellite network has some drawbacks \cite{Han2016b}. IETF's IP-based location management techniques were designed to manage the logical location of MNs (terminals) and deliver their  data to wherever they move. However, in LEO SatNets both terminals and BS (satellites) are moving, which creates new challenges that cannot be fully addressed using existing IETF's location management techniques. In addition, IETF location management techniques are intended to work in centralized units that manage both control and data traffic (i.e., routing) \cite{Cordova2019}. As a result, IETF location management techniques have poor scalability and may create processing overload in core network devices. Moreover, even in terrestrial networks, such standards posed several problems because of their low granularity mobility management and suboptimized routing. What makes things more challenging is the characteristics of future LEO SatNets, such as the very frequent and rapid topology changes due to the fast LEO satellites speed, the very dense deployment of LEO satellites in the form of a network of mega-constellations, and the complete integration with aerial, terrestrial, and even deep space networks. In addition, future LEO SatNets will be utilized in highly populated areas where thousands or millions of heterogeneous user devices can communicate directly with the LEO satellite (without going through a gateway). Hence, future LEO SatNets will create unprecedented mobility scenarios that require innovative solutions.  

To overcome the limitations of IETF IP-based location management, one of three approaches was followed by the existing studies on LEO SatNets location management. The first approach attempted to enhance or extend the IETF IP-based location management techniques \cite{Zhang2019, Dai2020}. The second approach is based on the split of the two roles of IP addresses (i.e., locator/identifier split) \cite{Feng2016, Feng2017}. The third approach focuses on utilizing \acrfull{sdn} for the purpose of topology (location) management \cite{Xu2018b, Shuai2018}. However, the existing studies are not considering the unique characteristics of future LEO SatNets, and the adoption of the existing solutions -as they are- will not be adequate. \textbf{By pointing out the advantages that existing studies have and what is required for future LEO SatNets, this study aims to shape the future of the needed and foreseen location management.}


\subsection{Existing Surveys and Tutorials}

A number of excellent surveys and tutorials related to mobility management were published over the past several years, involving discussions on the IETF's mobility management protocols and their proposed enhancements, reviews on the available location/identifier split architectures and mapping systems, and various surveys on the exploitation of software defined networks concepts for mobility management purposes. 

A comprehensive tutorial on mobility management in data networks was introduced in \cite{Bolla2014}. In \cite{JAIN2020}, the authors discussed the suitability of the existing mobility management solutions introduced by the standardization bodies (e.g.,  IEEE,  IETF,  \acrshort{3gpp},  and ITU) to be applied to 5G and beyond networks. \cite{Nikander2010} introduced a review on the architectures, design, benefits, and potential drawbacks of the \acrfull{hip} which is an inter-networking architecture and an associated set of protocols, developed at the IETF. The mobility management services in mobile networks were surveyed in \cite{ALSURMI2012}. For networks with self-organizing and self-configuring characteristics, the applicability of IETF mobility management protocols was discussed in \cite{Hasan2012}. Several algorithms, developed to address the challenges of IP-based mobility management in the \acrfull{iot} environment, were reviewed in \cite{Safwan2016}.

A detailed discussion of the limitations of the IP addressing architecture and the existing enhancements based on location/identifier split architectures were presented in \cite{Ramirez2014}. In \cite{Feng2017}, the authors introduced a comprehensive  survey  on  location/identifier  split  network  architectures  and  their characteristics. As an essential component in location/identifier split solutions, \cite{Hoefling2013} provided a survey on several mapping systems.

SDN is considered a promising approach to manage mobility. The author in  \cite{Elsadek2017}, discussed the challenges faced when SDN is utilized to manage mobility in IP-based networks. 

Mobility management is a quite mature research topic in communication networks, however, this is not the case for the next generation of SatNets. Many recent surveys and tutorial on future SatNets focused on discussing communication and networking related issues. For example, Radhakrishnan \textit{et al.} \cite{Radhakrishnan2016} focused on inter-satellite communications in   small  satellite  constellations from the perspectives of physical to network layers, and the Internet of remote things applications of satellite communication were reviewed in \cite{Sanctis2015}. However, only a few reviews were published on the mobility management related issues in next generation satellite networks. The author in \cite{Miao2016} discussed the challenges facing SDN-based integrated satellite-terrestrial networks. Another review \cite{Xu2018}, explored the challenges that software-defined next generation satellite networks may encounter and provided some potential solutions.  \cite{Hossain2018} discussed the survivability and scalability of space networks. The aforementioned surveys with regards to mobility management are summarized at a glance in Table \ref{SONSENComparison} to allow the reader to capture the focal point of each of the existing surveys.

The discussion throughout this section reveals that mobility management in future SatNets is still in its infancy phase. Although some very few reviews discussed the integration of SDN and future SatNets, mobility management issues, and more specifically location management, were not the focus of such papers. Therefore, this review paper aims to discuss the  existing location management solutions and the challenges facing their applicability in future LEO SatNets.

\begin{table*}[t]
\caption{Recent surveys and tutorials related to mobility management.} 
\centering 
\begin{tabular}{|p{0.7cm}|p{1.5cm}|p{14cm}|}
\hline
\cellcolor{aliceblue}\textbf{Year } &  \cellcolor{aliceblue}\textbf{ Publication }	 & \cellcolor{aliceblue}\textbf{ One-sentence summary} \\ 
\hline 
\hline

2010 & \cite{Nikander2010} & An in-depth look at the architecture, design, benefits, and potential drawbacks of the HIP. \\
 \hline
 2012 & \cite{ALSURMI2012} & A survey of the mobility management services and their techniques, strategies and protocols. \\
 \hline
 2012 & \cite{Hasan2012} & Discusses some future network architectures in terms of their support for IETF mobility management protocols, including architectures of wireless mesh networks with self organizing and self configuring network characteristics.\\
 \hline
 2013 & \cite{Hoefling2013} & A survey of mapping systems used in location/identifier split proposals\\
 \hline
 2014 & \cite{Bolla2014} & A comprehensive tutorial on mobility management in data networks.\\
\hline
2014 & \cite{Ramirez2014} & Discusses the limitations of the IP addressing architecture and reviews the existing enhancement proposals based on location/identifier split architectures. \\
\hline
2016 & \cite{Safwan2016} & A review of the developed algorithms to address the challenges of integrating IoT and IP, the attributes of IP mobility management protocols and their enhancements. \\
\hline
2016 & \cite{Miao2016} & A survey of the recent research works related to the software defined integrated satellite and terrestrial networks with challenge identification and 
a discussion of the emerging topics requiring further research. \\
\hline
2017 &  \cite{Elsadek2017} & A review of the studies on IP mobility management using software defined networking.\\
\hline
2017 & \cite{Feng2017} & A comprehensive survey on location/identifier split network architectures and their mechanisms, and characteristics.\\
\hline
2018 & \cite{Xu2018} & An architecture of software-defined next-generation satellite networks with the exploitation of network function virtualization, network virtualization, and software-defined radio concepts. \\
\hline
2018 & \cite{Hossain2018} & A survey on survivability and scalability of space networks with a discussion on IP mobility management applicability \\
\hline
2020 & \cite{JAIN2020} & A review of the 5G and beyond mobility management functional requirements with a discussion on whether the existing mechanisms introduced by standardization bodies (e.g., IEEE, IETF, \acrshort{3gpp}, and ITU) meet these requirements.\\
\hline
\end{tabular}
\label{SONSENComparison} 
\end{table*}

\subsection{Paper Contributions and Structure}

The contributions of this paper are as follows:
\begin{itemize}
    \item We describe the future LEO SatNets mobility characteristics and its challenging features and highlight two unprecedented location management scenarios. 
    \item We give an overview of IETF's location management techniques and their limitations in the context of future LEO SatNets. 
    \item We comprehensively and critically review the existing solutions for location management in LEO SatNets, which are categorized into three approaches.
    \item For each of the reviewed approaches, we point out the  \textit{``Issues to consider"} which are important points that should be considered \textcolor{black}{for that specific approach of location management to serve future LEO SatNets.}
    \item We summarize the current view of LEO SatNets location management. 
    \item Important future research directions are highlighted, including considering the relationship between orbit related parameters and location management, utilization of Blockchain technology, adoption of the collaborative Internet architecture, investigating the new IP address proposal.   
\end{itemize}

Section \ref{CharaAndChallenge} presents the future LEO SatNets envisioned mobility characteristics,  identifies the challenging features of future LEO SatNets, and describes two unprecedented mobility scenarios. In Section \ref{IPv6MM}, we give an overview of IP-based standardized location management and its limitations in future LEO SatNets.  Section \ref{taxonomy} introduces the taxonomy of the existing location management techniques for LEO SatNets. Section \ref{Approach1} reviews the existing studies that proposed extensions of IETF location management techniques for LEO SatNets. Section \ref{Approach2} discusses the existing solutions that followed the locator/identifier split approach in LEO SatNets. Section \ref{Approach3} investigates the SDN-based location management in LEO SatNets. At the end of Sections \ref{Approach1}, \ref{Approach2}, and \ref{Approach3}, an \textit{``Issues to consider"} subsection is included, which highlights the main points that should be taken into consideration for future LEO SatNets location management. 
Section \ref{Currentview} highlights the advantages of existing solutions of the three location management approaches in future LEO SatNets, and suggests important future research directions. Our conclusions are presented in Section~\ref{con}.



\section{Mobility characteristics in future LEO SatNets}\label{CharaAndChallenge}

Satellite communication systems have been considered as a potential solution for complementing terrestrial networks by providing  coverage in rural areas as well as
offloading and balancing data traffic in densely populated areas \cite{Papa2018}. With the emergence of LEO satellite mega-constellations, which involve hundreds to thousands of satellites \cite{Chaudhry2020} \cite{Chaudhry2021}, the concept of satellite networks is evolving rapidly and gaining increased attention. 
\acrshort{3gpp} introduced a number of satellite use cases in 5G networks (\acrshort{3gpp} TR 22.822 Release 16) which discuss the role of satellites in future networks \cite{3GPPRel162018}. For example, \acrshort{3gpp} introduced \textit{ Internet of Things with a Satellite Network} and \textit{ Global Satellite Overlay} use cases that both emphasize the future role of satellite networks. However, to realize such use cases there are still several challenging matters to address. A major challenge that faces future satellite networks is mobility management. Therefore, in this section, after highlighting the challenging features of future LEO SatNets, we discuss two unprecedented  mobility management scenarios in future satellite networks with a focus on location management.

\subsection{Challenging Features of Future LEO SatNets}
There are some key points that should be kept in mind while designing, implementing, or evaluating location management solutions for future LEO SatNets:
\begin{enumerate}
    \item A network of mega-constellations that have thousands of LEO satellites operated by different operators is expected. Therefore, it is essential to ensure interoperability between different constellations and also between the different operators. This necessitates the development of standards for future satellite network operation and management.
    \item LEO satellites move at very high speed which results in frequent handovers.
    \item Future LEO SatNets is not only to provide coverage for rural or remote areas but also to serve highly populated areas by boosting terrestrial network capacity and ensuring continuous coverage for fast-moving users (e.g., trains, planes, drones). Thus, thousands of users can be connected to a LEO satellite. 
    \item With the development of wireless communication technologies, not only large terminals but also small or handheld user devices for broadband communication will be able to communicate directly with satellites without the need for ground gateways. This will open the door of heterogeneity in terms of the used devices and their required \acrfull{qos}.  
    \item Future LEO SatNets will be integrated with terrestrial, aerial, and maybe deep space networks. In terms of network management, such integration will result in more complexity and require high scalability. 
    \item A satellite will have multiple roles as it can work as a terminal, a router, and a BS. 
    \item Although deploying satellites at low/very low altitudes will require hundreds or thousands of satellites to provide continuous coverage all over the globe, a significant decrease in propagation delays can be achieved in comparison to legacy satellite communication. Nevertheless, the resulting delay and jitter are still considered non-negligible in certain delay-sensitive applications.
    \item Deploying LEO satellites on low earth orbits will decrease the communication delay with terrestrial or aerial networks or users. However, the lower the orbit the more satellites are required to provide coverage, the faster the satellites should move, and the smaller the satellite footprint will be. Consequently, the frequency of handover will increase when orbit altitude decreases. 
\end{enumerate}

\subsection{Unprecedented Location Management Scenarios}

In future broadband satellite networks, satellites will no longer be only used as a bent pipe. A satellite will be working as a mobile BS, a router, and a terminal. However, a LEO-based mobile BS moves very fast, which results in a high frequency of handovers and location updates triggers for both the satellite and its connected users. Moreover, future satellite networks will have thousands of LEO satellites, which require location management solutions with high scalability. 
Due to the LEO satellite connectivity and mobility characteristics, future LEO SatNets will introduce two unprecedented location management scenarios that require new solutions. The following two points elaborate on the two envisioned scenarios:

\begin{figure}[h]
\centering
\includegraphics[width=0.5\textwidth]{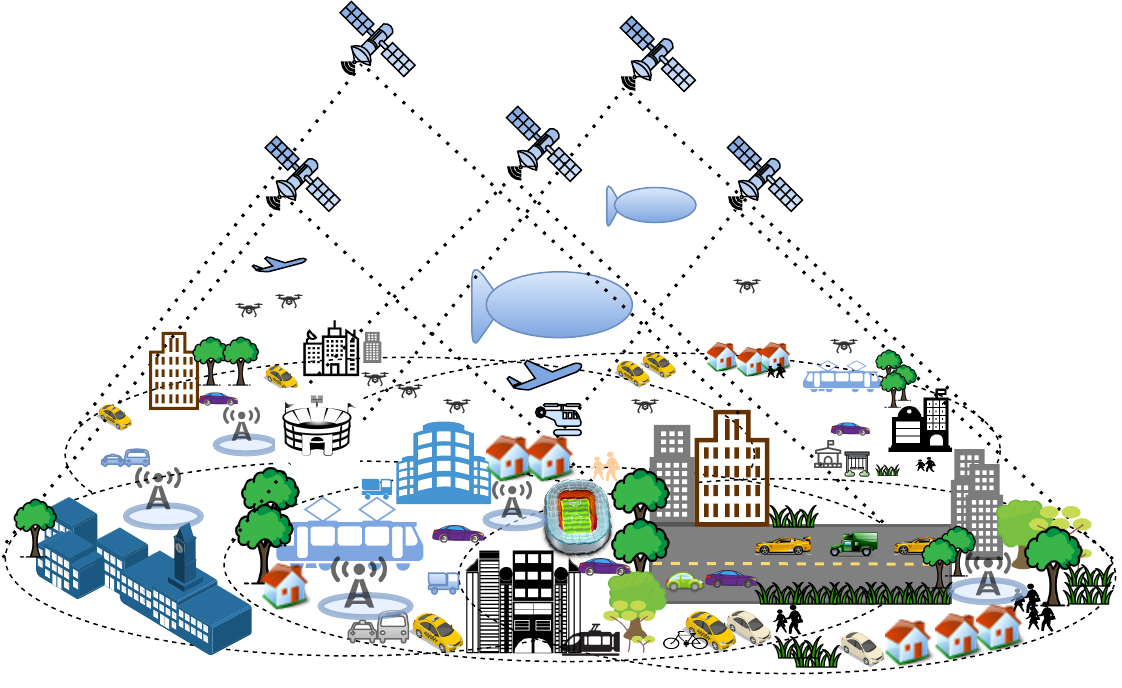}
\caption{LEO satellite based mobile BS serving thousands of users.}
\label{fig:FirstScenario}
\end{figure}

\begin{figure}[h]
\centering
\includegraphics[width=0.5\textwidth]{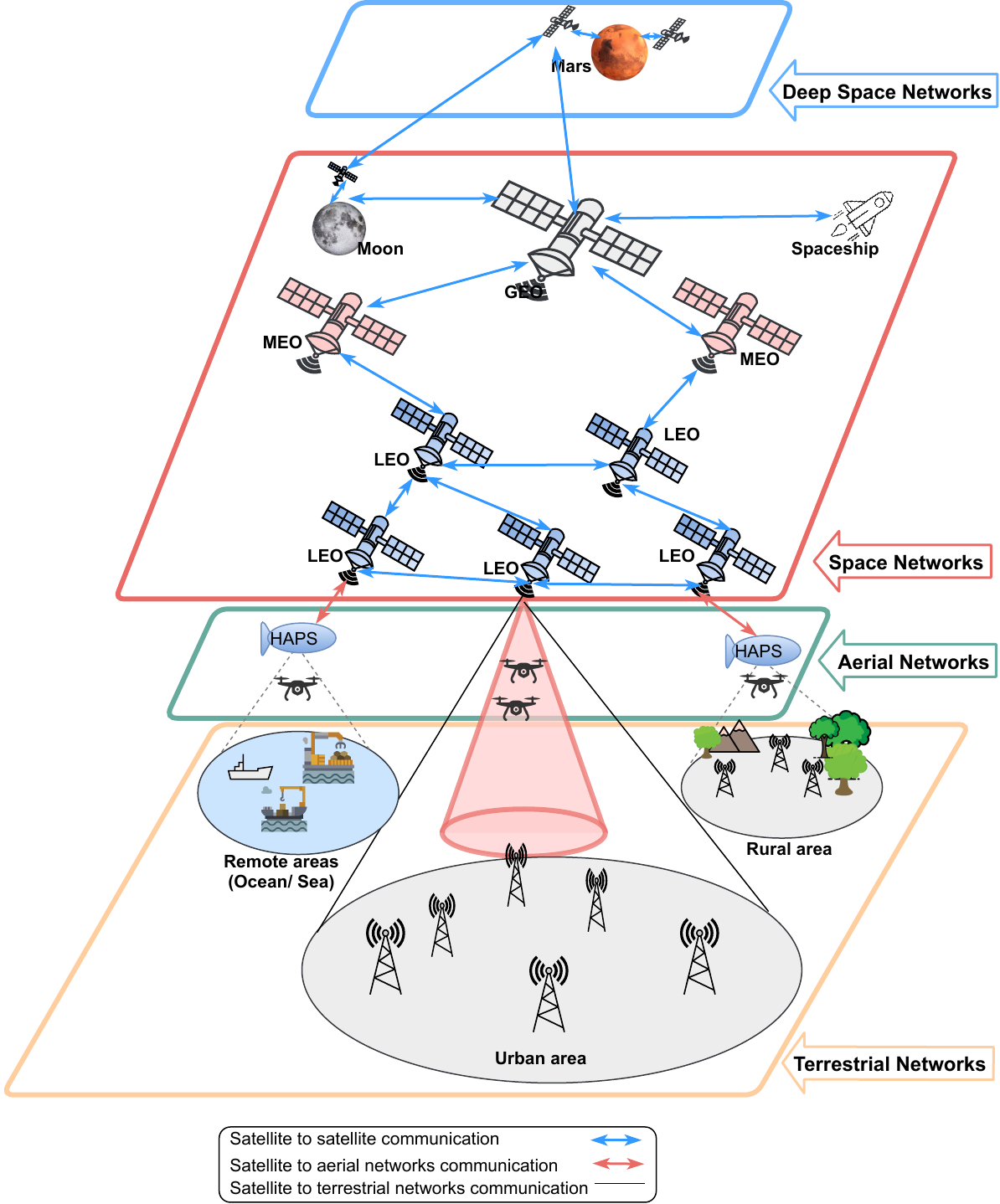}
\caption{LEO satellites connected to multiple networks.}
\label{fig:SecondScenario}
\end{figure}

\begin{itemize}
    \item \textbf{LEO satellite-based mobile BSs moving at high speeds and serving thousands of users' devices:}  In future networks, it is expected that satellites, especially LEO satellites, will provide wide coverage and support the communication network capacity in densely  populated areas, as shown in Figure \ref{fig:FirstScenario}. In this situation, a LEO satellite-based mobile BS will be serving  thousands of users. This will be empowered by the integration of reconfigurable intelligent surfaces with LEO satellites as well \cite{Tekbiyik2020}. In this scenario, a wide range of user device types can be served through LEO satellites including smart devices, machines, sensors, autonomous vehicles, and cargo drones. The mobility of LEO at high speeds will result in triggering location updates not only to the satellite based mobile BS but also to the thousands of users that are connected to the LEO satellite.  Although the users might not be moving, changing their network access point (i.e., the LEO satellite mobile BS) will trigger location updates in classical mobility management protocols (e.g., MIPv6). This is because, in IP networks, IP addresses are used for both routing and addressing purposes. Moreover, this unnecessary and massive number of location updates will be triggered every 5-10 minutes approximately. 
    
    \item \textbf{A LEO satellite can be connected to two or more networks simultaneously (i.e., terrestrial, aerial, and space networks):} In future integrated networks, besides being part of the network of satellite mega-constellations, LEO satellites will be also connected to terrestrial networks, aerial networks, or both, as shown in Figure \ref{fig:SecondScenario}. When a LEO satellite is connected to terrestrial and/or aerial networks, changes in satellite position will trigger location updates in both networks if each network has its own location management system. In addition, the topology changes in LEO satellite mega-constellations will also trigger frequent location updates among satellites. In this scenario, the LEO satellite can play the role of a mobile BS, a router, or a terminal. However, the most complicated case is when the LEO satellite is working as a mobile BS to serve a large number of users through multiple backhaul connections (space, aerial, terrestrial). Managing the location updates in several networks and providing a mapping between the location systems of such networks is considered a challenging issue. 
\end{itemize}

\section{Overview of IP-based standardized location management and its limitations in future LEO SatNets}\label{IPv6MM}

In traditional terrestrial cellular networks, mobility management is quite a mature topic. For location management, most of the research has focused on tracking and paging. \textcolor{black}{Tracking in cellular networks is the process of identifying in which cell the user (i.e., MN) is located while it is stationary or mobile through using the user's signal strength received by nearby cellular towers. Paging is the process of indicating a user position in the cellular network in order to establish a connection with another user calling from a fixed or mobile equipment.} The tracking area (location area) used in 4G and 5G usually comprises a dynamic group of cells. Location management solutions aim to find the balance among tracking area division and location updates/paging overhead. To communicate with other devices in a cellular network, the MN device must establish an end-to-end user plane path through the domain of the mobile operator. To manage the location of an idle MN, the MN  performs a location update upon crossing the boundaries of a tracking area. This location update is saved in a database that can be enquired to know the location of an idle MN. To discover an idle MN current location, the location area's cells  are contacted through paging. With the upcoming densification of 5G, it is expected to encounter a considerable increase in the  location management signaling cost due to more location updates (if the tracking areas are small) or paging (if the tracking areas are large) \cite{Kaloxylos2020}. 

In IP-based networks, location management is done in a slightly different way as the active TCP/IP connections of a MN need to be maintained while moving from one access router to another. In the 1970s, the IP protocol was introduced as an inter-networking protocol for delivering data packets in wired networks, where IP addresses play the double role of routing and addressing. With the development of mobile wireless communication devices, there was a real need to support mobility in IP-based networks. Therefore, IETF introduced MIPv4 and later followed by MIPv6, PMIPv6, \acrfull{fmipv6}, and \acrfull{hmipv6}. Mobility management, in these protocols, consists of two main components, which are handover management and location management. In this study, the focus is on the location management aspect of mobility management. Location management aims to locate the MNs and guarantee data delivery \cite{Shahriar2008}. The two procedures composing location management are binding updates and data delivery. To address mobility in Internet networks, the IETF mobile internet protocols bind the MNs to their corresponding new IP addresses as the MNs’ locations change. \textcolor{black}{A binding update is performed only when a handover has occurred (i.e., when the MN changes its network access point).} The following two subsections explore the fundamental procedure of location management in IPv6 mobility management standards and investigate the limitations of applying such standards in future LEO SatNets. Figures \ref{fig:MIPv6}, \ref{fig:FMIPv6}, \ref{fig:HMIPv6}, and \ref{fig:PMIPv6} describe the network architecture of each of the four IPv6 mobility management standards, and summarize their location management procedures.

\subsection{Location Management Procedure in IPv6 Mobility Management Standards}

\paragraph{\textbf{MIPv6 \cite{MIP2004}}} \textcolor{black}{In the home network (i.e., where the MN is currently located and attached) the MN gets a permanent address, called the home address.} This address is registered at the \acrfull{ha} in the home network and is used for both identification and routing purposes. Figure \ref{fig:MIPv6}, shows the network architecture and the location management message exchange of MIPv6. Since MIPv6 is a host-based mobility management protocol, the MN detects its mobility from the home network (previous network) to a foreign network by using the IPv6 neighbour discovery mechanism. \textcolor{black}{A foreign network is a network that the MN access after moving out of its home network coverage.} When the MN moves out of the home network and accesses a foreign network, it will perform the following steps \cite{Chen2016}:

\begin{figure}[h]
\centering
\includegraphics[width=0.5\textwidth]{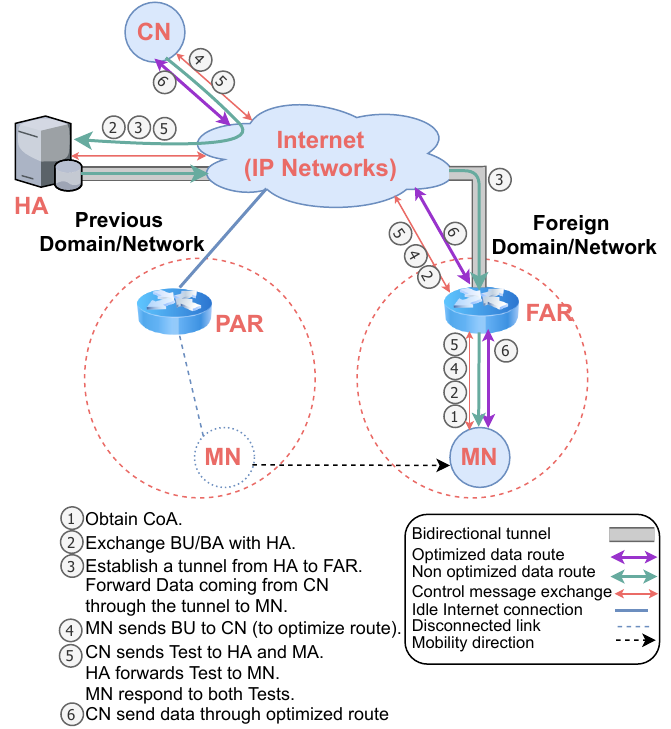}
\caption{MIPv6 location management procedure \cite{Chen2016}.}
\label{fig:MIPv6}
\end{figure}

\begin{itemize}
    \item The IPv6 neighbor discovery or address auto-configuration mechanism is used to obtain a temporary IP address from the foreign network, called the \acrfull{coa} \textcolor{black}{\textbf{(Step 1)}}.
    \item The MN informs the HA of its current location by sending a \acrfull{bu} message and the HA responds with a \acrfull{ba} to the MN \textcolor{black}{\textbf{(Step~2)}}.
    \item After completing the binding update with the HA, the HA and the \acrfull{ar} at the foreign network (i.e., \acrfull{far}) will establish a bidirectional tunnel to deliver the data packets between the \acrfull{cn} and MN \textcolor{black}{\textbf{(Step 3)}}.  In this case, the data packets have to traverse the HA, which is not necessarily the optimum route \cite{Zhang2019}.
    \item The MN has the option to optimize the data forwarding route by sending BU message to the CN as well. Nevertheless, the MN will keep receiving packets through the HA until the CN starts using the CoA address \textcolor{black}{\textbf{(Step 4)}}.
    \item Before using the CoA, the CN will send two test messages to both the HA and MN. The HA has to forward the message to the MN then the MN has to respond to both messages through the two different paths. When the CN receives both responses it can start using the CoA, then the communication between CN and MN can be through the FAR without going through the HA \textcolor{black}{\textbf{(Step 5, 6)}}.
\end{itemize}
    In MIPv6, the handover and the location management processes are closely coupled, and every handover results in updating the CoA at HA and CN (for route optimization). This leads to a high handover delay and increases packet loss rate. Therefore, some improved protocols, such as FMIPv6 \cite{FMIP2005}, HMIPv6 \cite{HMIP2005} and PMIPv6 \cite{PMIP2008}, have been proposed.

\paragraph{\textbf{FMIPv6 \cite{FMIP2005}}}
To reduce packet loss and  handover latency, IETF proposed FMIPv6,  which enables the MN to configure a CoA  before moving to the new AR (i.e., FAR) coverage \cite{Shahriar2008}. FMIPv6 protocol allows a MN to request information about neighbouring ARs. There are two modes of FMIPv6, namely Predictive and Reactive handover \cite{Pieterse2012}. Figure \ref{fig:FMIPv6} shows an example of FMIPv6 handover, where the MN location is updated through the following steps:

\begin{figure}[h]
\centering
\includegraphics[width=0.5\textwidth]{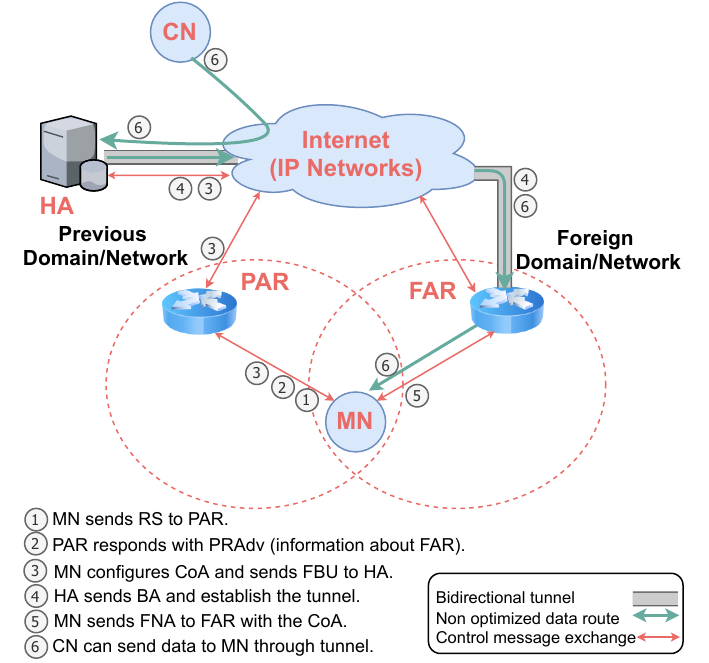}
\caption{FMIPv6 location management procedure \cite{Pieterse2012}.}
\label{fig:FMIPv6}
\end{figure}

 \begin{itemize}
     \item The MN sends a \acrfull{rs} to the \acrfull{par} requesting information for a potential handover \textcolor{black}{\textbf{(Step 1)}}. 
     \item PAR replies with a \acrfull{pradv} containing information about neighbouring ARs \textcolor{black}{\textbf{(Step 2)}}.
     \item After receiving PRAdv, the MN configures a CoA and sends a \acrfull{fbu} to FAR to bind the MN's home address to the CoA in order to tunnel the arriving packets to the new location of the MN. PAR sends an acknowledgement to the MN confirming that the tunnel is ready \textcolor{black}{\textbf{(Step 3, 4)}}. 
     \item The MN will send a \acrfull{fna} as soon as it gets connected to FAR. This message confirms the use of the CoA \textcolor{black}{\textbf{(Step 5)}}.
 \end{itemize}

However, if the MN could not anticipate the handover, then the reactive mode will be used. In this case, the MN will configure its CoA after moving to FAR coverage and the FBU is sent to PAR through FAR and is encapsulated in a FNA message. Then, the two routers PAR and FAR will exchange a handover initiation/handover acknowledgement messages to establish the tunnel then PAR starts forwarding packets to FAR to be delivered to the MN.

\paragraph{\textbf{HMIPv6 \cite{HMIP2005}}}

 HMIPv6 is an enhancement of Mobile IPv6 with the feature of localized mobility management for MNs  \cite{Shahriar2008}. To support localized mobility management, it introduces a new network entity called the \acrfull{map}. In HMIPv6 a MN has two types of addresses; a regional \acrfull{rcoa}  and an \acrfull{lcoa}. The RCoA is a global address and specifies a particular domain of the Internet. LCoA is a local address within the domain.  When the MN moves between local networks inside a MAP domain (micro/intra-domain handover), it changes and updates its LCoA only at the MAP. However, moving from one MAP domain to a new MAP domain (macro/inter-domain handover), the MN has to change both addresses by registering a new local LCoA and a new RCoA at the new MAP. In this case, the new MAP registers the new RCoA to the MN’s HA. Figure \ref{fig:HMIPv6} shows the network architecture of HMIPv6 and an example of a MN moving from MAP1 domain to the MAP2 domain \cite{Yoo2009}: 
 
 \begin{figure}[h]
\centering
\includegraphics[width=0.5\textwidth]{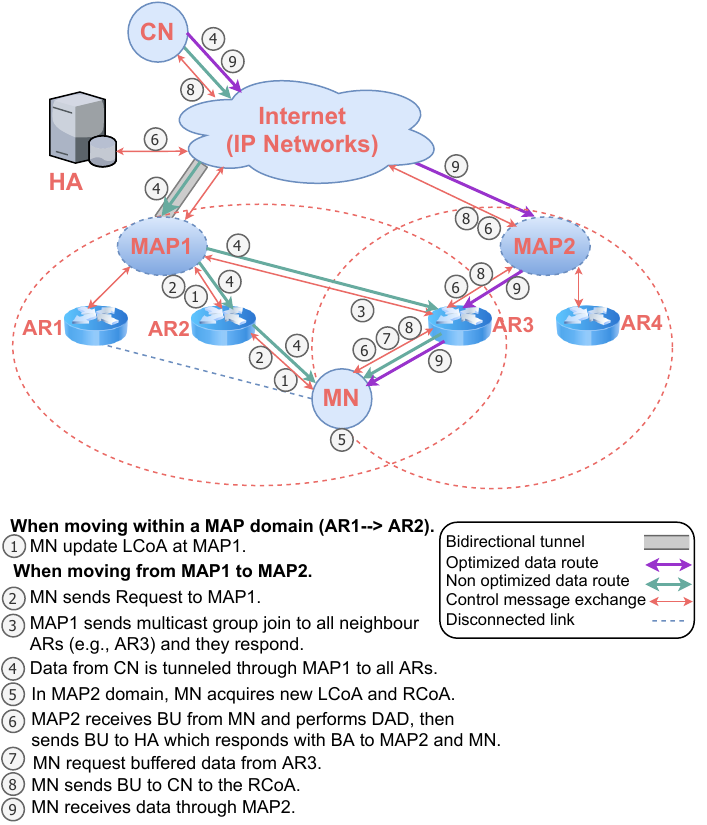}
\caption{HMIPv6 location management procedure \cite{Yoo2009}.}
\label{fig:HMIPv6}
\end{figure}
 
 \begin{itemize}
     \item The MN sends a request control message to MAP1 to create a multicast group for the MN \textcolor{black}{\textbf{(Step 2)}}.
     \item MAP1 creates a multicast group by sending a multicast group join request to all neighboring ARs. Then, the neighboring ARs respond to MAP1 to show their availability to receive multicast data packets \textcolor{black}{\textbf{(Step 3)}}.
     \item During the handover process, any received data packet from the CN is tunneled through MAP1 to all the available ARs, where it is buffered \textcolor{black}{\textbf{(Step 4)}}.  
     \item When the MN travels from MAP1 domain to MAP2 domain, it acquires new addresses (i.e., new RCoA, new LCoA) from the MAP2 network \textcolor{black}{\textbf{(Step 5)}}.
     \item The MN sends a BU to MAP2 through AR3 and sends a message requesting AR3 to forward a multicast message. AR3 receives the request message, and subsequently forwards the buffered packets to the MN \textcolor{black}{\textbf{(Step 7)}}.
     \item MAP2 receives the BU message and performs \acrfull{dad}. MAP2 sends a BU to the MN’s HA after receiving the DAD and waits for a BA from the HA. After receiving a BA from HA, MAP2 sends a BA to the MN \textcolor{black}{\textbf{(Step 6)}}.
     \item After receiving a BA, the MN sends a BU to the CN via MAP2 to change the destination address to the new RCoA. Then, data packets will be delivered to the MN through MAP2 \textcolor{black}{\textbf{(Step 8, 9)}}.
 \end{itemize}

\begin{figure}[h]
\centering
\includegraphics[width=0.4\textwidth]{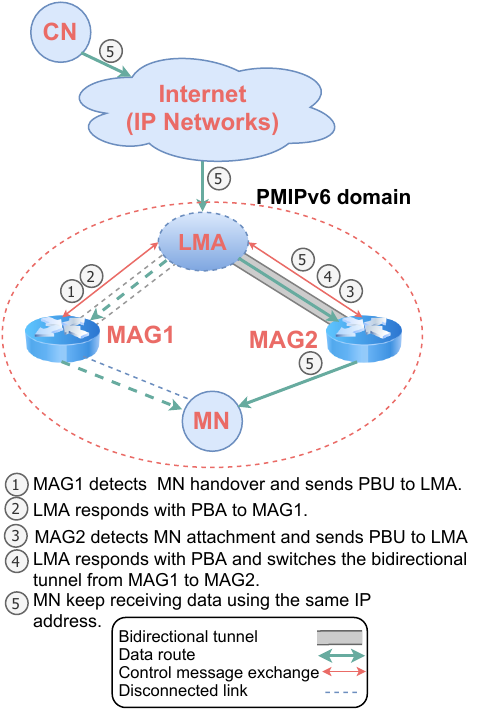}
\caption{PMIPv6 location management procedure \cite{Chen2016}.}
\label{fig:PMIPv6}
\end{figure}

\paragraph{\textbf{PMIPv6 \cite{PMIP2008}}}To provide a mobility management solution with reduced signaling and delay to support a MN moving within an IPv6 domain, the IETF introduced PMIPv6. As a network-based mobility management protocol, PMIPv6 introduces two new network entities, \acrfull{mag} and \acrfull{lma} \cite{Hussain2020}. A LMA is connected to multiple MAGs and in one PMIPv6 domain, there can be multiple LMAs managing the mobility of a different group of MNs. When the MN is moving within a PMIPv6 domain, the MAG performs the signaling interaction with the LMA on behalf of the MN to ensure session continuity \cite{Han2016b}. When a MN joins the network, it will send a RS to the reachable MAG. Then the MAG sends a \acrfull{pbu} to its LMA. The LMA responds with a \acrfull{pba} which includes the MN's home network prefix,  creates a \acrfull{bce}, and establishes a bidirectional tunnel with the MAG. The MN will use the home network prefix to configure its address using either stateless or stateful address configuration. When the MN is moving from the coverage of one MAG to another within the same PMIPv6 domain, as described in Figure \ref{fig:PMIPv6}, only a local update of location is required and data flow can directly be adjusted at LMA based on the following steps \cite{Chen2016}:

\begin{itemize}
    \item MAG1 detects that the MN is moving away from its coverage area and sends a PBU to the LMA \textcolor{black}{\textbf{(Step 1)}}.
    \item The LMA responds with a PBA message to MAG1~\textcolor{black}{\textbf{(Step~2)}}.
    \item MAG2 detects the attachment of the MN and sends a PBU to the LMA \textcolor{black}{\textbf{(Step 3)}}.
    \item The LMA responds with PBA message to MAG2 and switches the bidirectional tunnel from MAG1 to MAG2 \textcolor{black}{\textbf{(Step 4)}}.
    \item The MN keeps using the same IP address as long as it is moving between MAGs belong to the same PMIPv6 domain \textcolor{black}{\textbf{(Step 5)}}.
\end{itemize}

In case the MN moves outside the PMIPv6 domain, then the location management procedure of MIPv6 needs to be executed and the home network LMA will play the role of a HA.

\subsection{Limitations of IPv6 Location Management Standards in Future LEO SatNets}

 Location management goal is to locate mobile nodes and guarantee their data delivery while moving from one access point/network to another. The two phases of location management are binding updates and data delivery \cite{Shahriar2008}. \textcolor{black}{Unlike terrestrial networks that are geographically bounded, future SatNets may operate globally. This characteristic makes the adoption of the IETF’s mobility management standards infeasible. This is because the IETF's mobility management standards depend on the availability of fixed anchor nodes that manage MNs mobility in a centralized manner.}  Moreover, in all IP-based location management protocols,  data packet routing goes through the location management anchor thereby
producing the non-optimal routing path \cite{Han2016b}. This non-optimal data routing  is unacceptable in future LEO SatNets as it will consume link resources and increase delivery delays. The following paragraphs discuss the drawbacks of applying each of the main IP mobility management protocols (i.e., MIPv6, PMIPv6, HMIPv6, and FMIPv6) in future LEO SatNets with a focus on the protocols' location management aspect.  
 
 \paragraph{\textbf{MIPv6}} Both MIPv4 and MIPv6 protocols have been designed to manage mobility in Internet networks by binding the MN to its corresponding new address as its location changes. However, implementing MIPv6 in future LEO SatNets will face the challenge of satellites' high mobility that will generate a large number of binding update requests from both LEO satellites and their connected end-users, which consume a massive amount of network resources. In MIPv6 (as in MIPv4), data packets transmission is disrupted  during the handover period (i.e., handover latency). This latency comprises the required time to detect the movement, configure a new address, and to update the MN location \cite{Shahriar2008}. Packets sent to the MN during the handover period might be lost. In future LEO SatNets, depending on the received signal strength to detect mobility might be inaccurate because of the signal fluctuation, which may result in unnecessary address configuration and location update requests. The effect of atmospheric disturbances (e.g., rain) on the received signal strength should be taken in consideration. In addition, the propagation delay will prolong the time of the new address configuration and location updates especially if the mobility control entity is located on Earth. Although MIPv6 introduced the routing optimization to avoid keep sending the data packets through the HA, the non-optimal routing still cannot be neglected at the initial stage. This may require the data packets to go through the HA, which can be a ground gateway, and then be sent to the destination satellite. With the long propagation delay of the \acrfull{gsl}, sending data packets down to Earth and then up  to satellites might create serious packet delivery delays and increase the load on  GSLs. This issue gets even worse when the HA is located far away from the current location of the satellite. \cite{Han2016b}.

\paragraph{\textbf{FMIPv6}} Under the FMIPv6 framework, the next access point is predicted and the address configuration of the MN can be done prior to handover to reduce the handover delay \cite{Shahriar2008}. However, FMIPv6 introduces some interactive signaling messages between the current and the new access routers and also requires the establishment of a tunnel between the two routers. Although FMIPv6 with buffering and forwarding mechanisms outperforms MIPv6 in reducing handover latency and packet loss, this comes with a cost. Basically, the forwarding tunnel between the current and new access routers is established prior to handover, and the sent data from
CN to MN is forwarded through the current router to the new one \cite{Su2017}. In future LEO SatNets, if satellites are playing the role of access routers, then creating a forwarding tunnel will consume bandwidth resources of \acrfull{isl} and satellite buffering capacity. In addition, as FMIPv6 depends on predicting the handover target (next access router), inaccurate predictions will waste network resources. In the presence of multiple mega-constellations, the user will have multiple potential satellites as a handover target, which makes predicting the handover target accurately a challenging task. 

\paragraph{\textbf{HMIPv6}} The HMIPv6 protocol adds a MAP to the network to handle local handover, which decreases the required mobility management signaling and reduces the handover delay of location updates \cite{Shahriar2008}. However, the large scale movement of a LEO satellite is considered as global mobility that cannot take advantage of HMIPv6. Thus, with the large scale of future LEO SatNets, HMIPv6 will be performing inter-MAPs handovers which generates a high number of control messages for location management and prolongs latency.

\paragraph{\textbf{PMIPv6}} In PMIPv6 the network performs the mobility management process on behalf of the MN, which reduces the signaling interaction between the MN and the network access point \cite{Hussain2020}. In \cite{He2016}, the author compared the performance of MIPv6 and PMIPv6 in a simple LEO constellation and the results showed reduced handover latency with the implementation of PMIPv6. However, the application of PMIPv6 in future LEO SatNets may face several drawbacks, such as the high load on LMA, the long handover delay due to the signaling that needs to pass the MAG and LMA which one of them might be located on the ground.
In PMIPv6, LMA manages not only the mobility of the MN but also handles its related data traffic \cite{Han2016b}. In future LEO SatNets, if a terrestrial gateway is the candidate  LMA, directly applying PMIPv6 can cause non-optimal routing. This is because packets cannot be routed among satellites of different domains instead packets would make a round trip through GSLs which is unnecessary \cite{Han2016b}. PMIPv6 can provide good mobility support for receivers during an IP multicast session \cite{Jaff2015}. However, when the source node is mobile (i.e., LEO satellite) in a PMIPv6 based multicast session, all the receivers need to resubscribe every time the source node changes its network access point or location.

\section{Taxonomy of location management approaches in IP-based LEO SatNets }\label{taxonomy}

\begin{figure*}[h]
\centering
\includegraphics[width=0.8\textwidth]{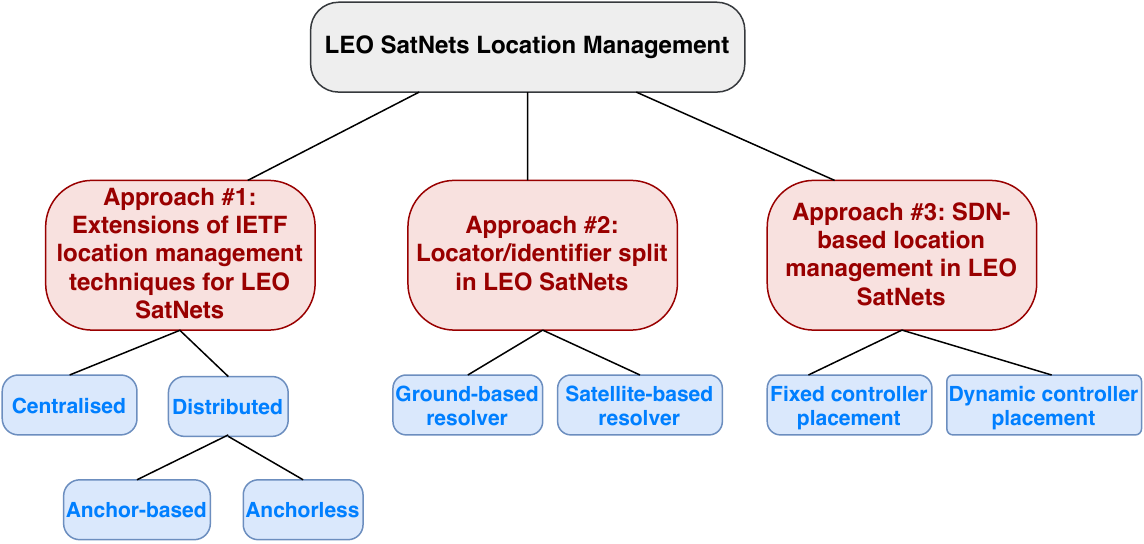}
\caption{Taxonomy of LEO SatNets location management.}
\label{fig:taxonomy}
\end{figure*}

The existing research about location management in LEO SatNets can be divided into three approaches, as described in Figure \ref{fig:taxonomy}:

\begin{enumerate}

    \item \textbf{Extensions of IETF location management techniques for LEO SatNets:} As described in Section \ref{IPv6MM}, mobility management in IP-based networks consists of handover management and location management. The IETF IPv6 mobility management standards (e.g., MIPv6, PMIPv6, FMIPv6, HMIPv6) addressed the location management issue in terrestrial networks. Although some research attempted to employ the location management techniques of IPv6 mobility management standards \cite{He2016}, \cite{Han2018}, \cite{Jaff2014}, such techniques have many limitations when applied to satellite networks. To enhance the performance of the IETF location management techniques, a number of extensions were proposed for satellite network location management, which are discussed in Section \ref{Approach1}. This solution's approach consists of two categories where the location management is done in either a distributed or centralized manner. The distributed IETF location management techniques' extensions can be either anchor-based or anchorless, as described in Figure \ref{fig:taxonomy}.
    
    \item \textbf{Locator/identifier split in LEO SatNets:} \textcolor{black}{The IP dual-role (i.e., locator and identifier) is regarded as the main cause of inefficient location management.} In terrestrial networks, many research works are investigating  the separation of the locator and identifier roles of IP such as \acrfull{ilnp} \cite{Atkinson2012}. Section \ref{Approach2}, explores the existing work on locator/identifier split in LEO SatNets and discusses the applicability of such solutions in future LEO SatNets.
    
    \item \textbf{SDN-based location management in LEO SatNets:} The SDN concept was introduced to add programmability and flexibility to network management \cite{Alaez2018}. Since the centralized nature of SDN limits the network scalability, several works have integrated SDN with a \acrfull{dmm} architecture to adapt to the large scale of LEO SatNets. Section \ref{Approach3} describes the existing studies that investigated the merging of SDN and DMM in LEO SatNets, and discusses the shortcomings of applying such solutions in future LEO SatNets location management. 
 
\end{enumerate}

The following three sections critically review and compare the existing studies under each approach and point out the important points that should be considered in the context of future LEO SatNets. In Section \ref{Currentview}, Table \ref{CompariosonTable} compares the advantages and the challenges of the three approaches from the perspective of future LEO SatNets.

\section{Approach \#1: Extensions of IETF location management techniques for LEO SatNets}\label{Approach1}

To provide continuous worldwide communication and Internet services, 
one of the main issues in future IP-based LEO SatNets is mobility management, which is more complex than in terrestrial networks because of the reasons mentioned in Section \ref{CharaAndChallenge}. From the perspective of future LEO SatNets, this section explores and discusses the proposed enhancements of existing IP-based location management standards and their drawbacks. This section is concluded with an ``\textit{Issues to Consider}" subsection, which describes some critical points that should be taken into consideration while implementing or developing IP-based location management solutions for future LEO SatNets.

\subsection{Enhancements of Existing IP-based Location Management Techniques for LEO SatNets}

SIGMA is a mobility management scheme where the MN can keep using its old IP address while obtaining the new IP address \cite{Shahriar2008}.  Every time the MN obtains a new address, it updates the \acrfull{lm} database and sets this new address as its primary address. To start a communication, the CN queries the LM with the MN's identity, and the LM replies with the primary IP address of the MN. Then, the CN can initiate communication with the MN in its new location. When dealing with satellites, the scheme uses satellite predicted mobility to predict the time of setting the primary address to the new IP address and delete the old IP address. However, the scheme did not consider the extensive signalling resulting from frequent satellite handovers in future mega-constellations. 

To decrease the location management cost, fixed Location Areas (LAs) can be chosen for location management in IP-based LEO SatNets. Fixed geographical location areas fulfill this requirement as it reduces the bending update frequency. However, when delivering data to a MN, huge and complex operations should be done by the network to determine to which satellites the MN is connected at that moment and to hide the effect of frequent satellite movement from the MN \cite{Zhang2019}. A location management scheme based on dual \acrshort{la}s in an IP-based LEO SatNet is proposed \cite{Zhang2012}. The scheme used two types of LAs, the \acrfull{fes} LA and satellite LA. Every FES is connected to three LEO satellites. Initially, MNs report both the satellite LA and FES LA information to its HA. A binding update will not be triggered unless the MN moves out of the two LAs that were reported to the HA through the last binding update procedure. Although this scheme suppresses the binding update frequency, its  loose location management necessitates  the use of paging to locate MNs (to which one of the three satellites the MN is connected). To send a packet to an ideal MN, the packet is first routed to the MN’s HA. Then the data packet is routed to the FES.  The FES sends a paging request to the satellite to which the MN has been registered (the last stored SAT ID). If the MN is still under the coverage area of that satellite, the packet will be delivered successfully. Otherwise, the FES predicts the satellite that covers the MN and sends the paging request to it. Clearly, this scheme is reducing the cost of binding updates while introducing the cost of paging and suboptimal data packet routing.

\begin{figure}[h]
\centering
\includegraphics[width=0.5\textwidth]{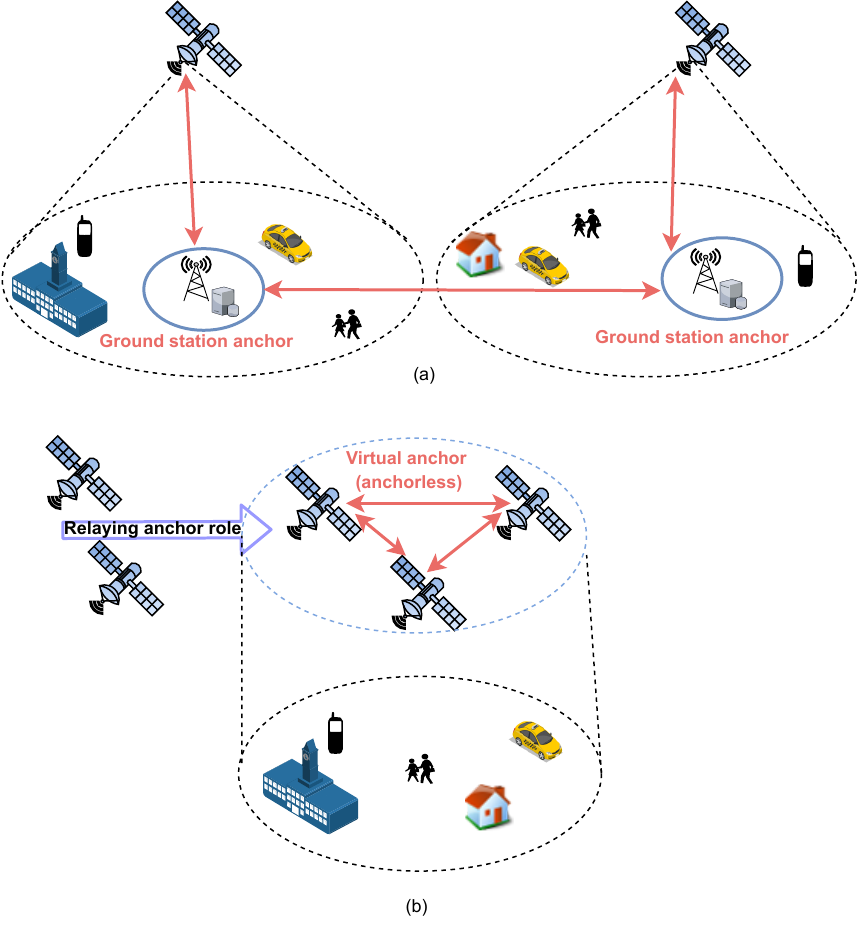}
\caption{(a) Anchor-based approach (using ground station anchor). (b) Anchorless approach (using a set of satellites as a virtual anchor for a certain LA).}
\label{AnchVsAnchless}
\end{figure}

To overcome the scalability issue of the centralized IP-based location management solutions, some studies proposed distributed solutions that come with the benefits of the optimal or near-optimal routing path, workload distribution, improved handover performance with shorter packet delivery latency. There are two types of distributed location  management, anchor-based and anchorless \cite{Jeon2017}. In anchor-based location management, the responsibilities of location management are permanently assigned to certain network entities. In contrast, anchorless location management role is shifted from one network entity to another based on network topology changes.  Figure \ref{AnchVsAnchless} compares the anchor-based and anchorless approaches. 

In \cite{Han2016}, the author presented an IP-based distributed location management scheme. The scheme is anchor-based as it depends on the availability of distributed \acrfull{gs} that are able to communicate and collaborate to manage the locations of satellites and attached MNs. The GSs register the binding (location) information of MNs and satellites, and they also forward the data from and to the MN. When a CN needs to communicate with a MN, it will forward the data through a satellite to the GS, then the GS will forward the data through satellites to the MN's corresponding GS. Thus, forwarding data packets has to go through GSLs which is considered a non-optimal route that consumes GSLs bandwidth and increases the packet delivery delay. Although, distributing location management tasks over GSs improves the system scalability, but it introduces a large amount of signaling overhead in the terrestrial network when binding updates are globally exchanged among GSs.

 A virtual mobility management scheme called VMIPv6, which is an enhancement of MIPv6 protocol, is proposed in \cite{Zhang2019}. VMIPv6 adopts the anchorless concept of location management and the distributed architecture introduced in the IETF's DMM requirements document (RFC 7333) \cite{Chan2014}. To reduce the location management overhead and delay, the author created a \acrfull{vac} to co-manage the mobility of users in the corresponding \acrfull{vad}. The set of LEO satellites on top of one specific LA is called VAC. The whole coverage area of all satellites in a VAC is defined as VAD. With changes in topology, the VAC is reconstructed  by  adding the new satellites sliding into the LA and deleting the ones sliding out of the LA. The  departing satellite relays the LA's binding information to the new satellite. Every VAC has  multiple local Mobile Agent Anchors (MAAs) and one \acrfull{hmaa}. The \acrshort{maa} and HMAA are on-board routers in LEO satellite to provide location management and routing services for the registered MNs. The MAAs of a VAC  share the mobility information of the MNs and cooperatively manage their binding. The HMAA maintains the connection between VAC and HA, and the MN registers its HMAA's subnet IP address at its HA. The local MAA is responsible for controlling the connection links between the MN and the VAC, and the MN binds its local MAA's IP address to each MAA of its related VAC. Within a VAD each MN has a global care-of address and a local one. When  satellite's mobility forces the MN to switch its connection to a new satellite within the same VAD, then the MN only updates the binding information with the new MAA. Thus, only the local care-of address will change. If the HMAA of a MN slid out of the VAD or the MN moved to another VAD, then the global care-of address will change. In this case, the MN should re-choose a MAA in VAC as its new HMAA and send the binding update information to the HA/CN to inform its new global care-of address.

The author of \cite{Dai2020} identifies two main drawbacks of placing the home agent entity in a ground station, which are: 1) ground stations are fixed and do not move with satellites, which makes it hard to communicate with the home agent when the satellite is not in line-of-sight; and 2) ground stations deployment is bounded by Earth geography. In addition, fixed home agents on satellites will require several hops to complete binding updates when the satellite is not in line-of-sight, which increases the update delay and consumes ISLs bandwidth. To overcome such problems, \cite{Dai2020} proposes to use a flexible agent placed on LEO satellites, where the home agent functionality is relayed from one satellite to another (i.e., the satellite that is closer to the MN) in a flexible manner. Once the procedure of binding update at the correspondent node is finished, the functionality of the home agent will relay from the previous flexible agent to the current access satellite. Although this solution reduces the delay in communicating with the home agent, frequently transferring the home agent records from one satellite to another will consume resources of ISLs. Nevertheless, having the HA on a satellite may result in forwarding the data packets through satellites even though the CN and the MN are connected through terrestrial networks. Table \ref{TableIPbased} shows a comparison of the IP-based location management enhancements for LEO SatNets, and highlights their limitations with respect to future LEO SatNets. 



\begin{table*}[t]
\caption{Comparison of  IP-based location management enhancements for LEO SatNets.} 
\centering 
\begin{tabular}{|p{1cm}|p{2.5cm}|p{2cm}|p{2cm}|p{3cm}|p{2cm}|p{3cm}|}
\hline
\cellcolor{aliceblue}\textbf{Algorithm} &  \cellcolor{aliceblue}\textbf{Location management placement}	 &
\cellcolor{aliceblue}\textbf{Anchor based/ Anchorless}	 & \cellcolor{aliceblue}\textbf{Distributed/ centralized} & \cellcolor{aliceblue}\textbf{Location update frequency} &
\cellcolor{aliceblue}\textbf{Data forwarding route} &
\cellcolor{aliceblue}\textbf{Main limitation in LEO satellites mega-constellation}\\ 
\hline 
\hline
SIGMA \cite{Shahriar2008} & Ground & Anchor-based & Centralized & Every time the MN moves from one AR (satellite) coverage to another & It has to pass through terrestrial gateways & Did not consider the high frequency of satellite handovers in future LEO SatNets. Non-optimized routing through terrestrial networks. \\
\hline
Dual LAs \cite{Zhang2012} & Ground & Anchor-based & Centralized & When the MN leaves both FES LA and satellite LA & It has to pass through HA and FES & Requires paging to locate MN. Did not consider satellite's location management. \\
\hline
Distributed location management using GSs \cite{Han2016} & Ground & Anchor-based & Distributed & Every time a MN/satellite changes its GS  & It has to pass through GSs & Assumes the availability of many GSs. Did not consider the direct communication between satellites and MN. \\
\hline
VMIPv6 \cite{Zhang2019}& Satellites (location management), Ground (home agent) & Anchorless & Distributed & When the MN switches its connection to a new satellite within the same VAD, the MN only updates the local care-of address with the new MAA. If the HMAA of a MN slid out of the VAD or the MN moved to another VAD, then the global care-of address will change at HA. & It has to pass through HA then satellites & When a VAD is serving thousands of MNs, relaying MNs' binding records from the departing satellite to the new satellite consumes ISLs resources. Having a fixed HA on the ground causes non-optimal routing. \\
\hline

Flexible agent \cite{Dai2020} & Satellite & Anchorless & Every time a MN changes its access satellite & Distributed & It has to pass through the HA which is the current access satellite & In future LEO SatNets it is not easy to determine the next access satellite since there will be multiple candidate satellites in the mega-constellations and LEO satellites have limited processing resources.   \\
\hline
\end{tabular}
\label{TableIPbased}
\end{table*}

\subsection{Issues to Consider}
IP-based location management has been early investigated in GEO satellite mesh networks \cite{Jaff2014}. Nevertheless, the emergence of future LEO SatNets creates the need for more recent research on location management solutions that consider the special characteristics of future mega-constellations. The following points summarize some important issues that should be considered in IP-based location management for future LEO SatNets.  
\begin{itemize}
    \item In IP-based location management, the anchor entity has two roles managing terminals locations and routing data packets to terminals. When anchor nodes are placed on earth, location management in future LEO SatNets faces the problem of limited link resources and long propagation delays while communicating with anchor nodes for binding updates or data delivery. On the other hand, placing anchor nodes on LEO satellites may encounter the problem of limited on-board processing and storage, and satellite fast mobility. 
    \item  Studying the placement of anchor nodes (in both anchor-based and anchorless solutions) is very important to achieve route optimization. Through inspecting the previous studies, it is clear that neither space placement nor terrestrial placement of anchor nodes can give favourable routing performance in all forwarding scenarios.   
    \item IP-based location management solutions require complex signaling, such as the tunnels dynamic construction and release, which increases the load on satellites' OBP units. 
    \item The proposed enhancements of IP-based location management in LEO SatNets face the problem of high location management overhead due to the unprecedented network architecture, where satellite mounted BSs move at high speeds causing the frequent handover of users/MNs in large groups.
    \item Unlike location management in terrestrial networks, IP-based location management in future LEO SatNets has two levels.  The first level is the location management of MNs. The second level is the location management of satellites that act as a BS, router, or terminal. Separating the two levels of location management might reduce the complexity of the location management system. However, there is still a need for some kind of mapping and coordination between the two levels.
\end{itemize}

\section{Approach \#2: Locator/identifier split in LEO SatNets}\label{Approach2}

The current satellite network architecture is using IP addresses as both identifiers (identify who is the endpoint) and locators (i.e., where is the endpoint in the routing system). However, the dual role of IP address is diminishing its ability to support mobility, especially in future SatNets where the support for mobility with high scalability and tight time constraints is a pressing requirement \cite{Han2016b}. Mobility support in IP networks depends heavily on the network topology that has static anchor nodes, which makes IP mobility solutions impractical when applied to satellite networks. In a satellite network, location management should be intrinsically designed with the consideration of the network topology in order to avoid scalability issues. Some emerging mobile network architectures (e.g., MobilityFirst \cite{Venkataramani2014}, \cite{Karimi2017}), considered the separation of identifier and location as a contributory to mobility management enhancement. In particular, the scalability of routing with the implementation of the locator/identifier split  has been well investigated \cite{Han2016b}. With locator/identifier splitting, a remote node can be identified even if it is using multiple addresses during the communication (e.g., multi-homing concept). Thus, with locator/identifier separation, it is possible to keep an ongoing communication continuous since  moving MNs can keep their identifiers \cite{Wang2012}. 

Conventional mobility management consists of two main procedures, location management, and handover management. However, in the location/identity split approach, mobility management is achieved through two correlated steps, namely location (binding) update and location resolution \cite{Zhang2017}.

Based on the location/identity split approach, HIP \cite{HIP2006}, \cite{HIP2017} was proposed as a draft at IETF in 1999 followed by subsequent improvements in various aspects. The HIP architecture adds a new layer,  called the Host Identity Layer, between the IP layer and the transport layer, thereby decoupling the layers from each other, and splitting the dual roles of IP addresses. When HIP is used, IP addresses function as pure locators. Instead of IP addresses, the applications use Host Identifiers to name peer hosts. To establish a HIP association, the two involved communicating parties issue a four-way handshake. HIP has a number of implementations such as OpenHIP \cite{OpenHIP} and HIPL \cite{HIPL}. However, the implementation of  HIP for satellite networks has not been investigated.   

 \acrfull{lisp} \cite{Farinacci2013} was initiated
at the \acrfull{irtf} in 2007, and has developed in the IETF working group since 2009. LISP has been promoted by Cisco and many research organizations such as LISP4.net \cite{LISP4}, and LISP-Lab \cite{LISPLab} with worldwide testbeds. LISP has multiple implementations such as Open-LISP \cite{Phung2014}, Cisco IOS \cite{LISPCisco}, which significantly accelerated its development. LISP divides the whole network into the core and the edges and divides the IP addressing space into the \acrfull{eid} and \acrfull{rloc}. An EID is topology-independent and used as a local address by hosts within edge networks, while a RLOC is used as a global locator to transmit packets within the core network. In LISP, the border router that forwards packets from the edge to the core is called \acrfull{itr}, whereas the one that forwards packets in the opposite direction is called \acrfull{etr}.  The ITR maintains a cache of RLOC-EID mapping locally. If the ITR does not have the location of the destined EID, it will send a Map-Request to the mapping system. Afterward, the ITR will send the data packet to the proper ETR. There are a number of proposals for LISP mapping systems such as LISP-Tree \cite{Jakab2010} and LISP-DDT \cite{Fuller2012}. However, the application of LISP was not investigated in satellite networks. 

In \cite{Han2016b}, GRIMM is proposed as a gateway based regional mobility management architecture for satellite network based on locator/identifier split. GRIMM divides the coverage of the satellite system into regions based on the terrestrial gateways distribution. Each gateway is equipped with a \acrfull{tgms} and is responsible for the localized location management of the MN within its region. The global location management is realized through the synchronization among all gateways. When a foreign TGMS receives global updates, it generates the mapping entry between MN's ID and its corresponding TGMS. To avoid the non-optimal routing, GRIMM's location resolution is conducted before forwarding data packets. In GRIMM, a MN accesses a satellite with a fixed \acrfull{id} and the local TGMS records the MN's ID and its corresponding accessed satellite. When the MN ID is registered locally for the first time, the TGMS will trigger a global update among all TGMSes. To start a session between a MN and a CN, the accessed satellite will first send a request of location resolution to the local TGMS upon receiving the first data packet. If the local TGMS does not have the specific locator of the enquired  CN's ID, the request will be redirected to the corresponding TGMS. After receiving the requested CN location, the source satellite begins to encapsulate the data packet with location information and then forwards it through inter-satellite links to that area. The destined satellite can decapsulate the packet and then sends it to the CN. When a MN moves from one satellite area to another, subsequent messages will notify the accessed satellite of CN to update MN's location in the cached mapping entry. This is to update the routing path between satellites. Although the regional location management greatly reduces the management cost and facilitates the scalability of the network, global updates may create high signaling overhead among the gateways. In addition, keeping records of every MN's ID and its corresponding TGMS in all foreign TGMS requires massive storage resources. 

SAT-GRD is a proposed identification/location split network architecture for integrating satellite and terrestrial networks \cite{Feng2016}. It separates the identity of both  the network and the host from their locations. It introduces a hierarchical mapping and resolution system that enables the separation of the control and data plane in SAT-GRD as well as the decoupling of the intra-domain routing policy  from the inter-domain routing. Each edge network has its own edge mapping system and there are two core mapping systems one in space (using GEO and MEO satellites) and the other one is on the ground (for terrestrial network). Between satellite and terrestrial networks, there are a number of border routers that handle the mapping of the host's ID and location between the two networks.

A heterogeneous satellite-terrestrial network architecture, HetNet, is proposed in \cite{Feng2017b}. HetNet merges locator/identity split and information-centric networking. The author assumed the availability of a network manager node at each edge network that handles the location registration and the location-to-ID resolution. In addition, there are network management nodes in the core network that forms a hierarchical structure with the edge network management nodes. For satellite networks, their network management nodes are placed in the ground gateways. When a MN moves within the same network then its location is updated in the local network management node, whereas an upper network management node needs to update the MN location only when it moves from one network to another. However, the author did not clarify whether a satellite moving from one gateway to another will cause a local or a higher level location update. Moreover, the proposed architecture did not consider the direct communication between satellites and MNs, instead, it assumed that MNs can communicate with satellites through ground gateways only.

 Location/identity split can enhance mobility in satellite networks. \textcolor{black}{ However, with the high mobility of LEO satellites, employing the conventional binding (location) update schemes will create a large number of binding updates for both MNs and satellites, each with a high binding update rate.} To mitigate the effect of frequent satellite handover on the binding update rate, the authors of \cite{Zhang2017} and \cite{Zhang2016b} proposed the concept of \acrfull{vap} to make binding update independent of satellite's motion, where the VAP stays in fixed position in relative to the ground. The VAP scheme decouples the binding of endpoints and satellites into two types of independent bindings: the binding between the endpoint and the VAP, and the binding between VAP and physical satellite. The two independent bindings provide the binding information required for endpoint mobility based on the location/identity split approach. Thus, a virtual spherical network consisting of fixed VAPs is superimposed over the physical satellite topology in order to hide the mobility of satellites from the terrestrial endpoints. A VAP is created and maintained by the  satellites that pass over the fixed network location of VAP. Then a binding between the MN identity and the fixed virtual attachment point rather than the  physical satellite passing over the MN is carefully maintained by an identity-to-location resolution system. For the binding of the physical satellites to a certain VAP, the proposed scheme takes advantage of the periodic and predictable LEO satellite movement as well as the satellite predefined constellation topology. Thus, on a periodical basis, a group of satellites will be leaving a VAP and rebinding to the next VAP.   

To enable location/identifier split in satellite networks, there is a real need for a rapid mapping system that can resolve identifiers to network locations in a real-time manner. Conventional ground station based satellite system control is restricted to the land distribution. The distributed mapping system formed only by ground stations may be too far to access for global scattered mobile users. Consequently, the \textcolor{black}{location resolution latency} becomes another challenge. To address this issue, \cite{Zhang2017b}  presented a space-based distributed \acrfull{rmrs} along with a dynamic replica placement algorithm.  The goal of RMRS is to achieve low \textcolor{black}{location resolution latency}, low update cost, and high system availability (resilience to failures). The two main components of RMRS are the virtual resolvers and replica placement controllers. The space-based fixed virtual resolvers are responsible for maintaining the mapping replicas and responding to user requests, while the replica placement controllers on the ground are responsible for determining the number and locations of the mapping replicas. The virtual resolvers have the same concept as the VAPs \cite{Zhang2017}. The virtual resolvers are fixed with respect to Earth and create a virtual overlay network upon underlying moving physical satellites. Each virtual resolver is maintained at a certain time by specific satellites. When a satellite leaves a virtual resolver region, then the mapping records handled by this virtual resolver are then transmitted to the subsequent satellites passing by. The replica placement controllers usually reside in the ground stations and communicate with the virtual resolvers (LEO satellites) through GSLs. Replicating every mapping at every possible location will create a high cost, especially with the rapid movement of satellites and their large-scale network. Therefore, the replica placement controllers use the dynamic replica placement algorithm to determine the number and locations of replicas for each mapping entry so as to provide a tradeoff between \textcolor{black}{location resolution latency} and update cost. However, an accurate calculation of the region size is required with consideration of the number of satellites in the constellation. This is to avoid the situation of having no satellite in the virtual resolver region for some time.
  

Locator/identifier split also has its natural
advantages in mobility support. The identifier uniquely represents the node in the network and the varying locator is used for routing. Through dynamic mapping from identifier to locator before sending packets, the independent location management is achieved, and non-optimal routing is mitigated. Independent implementation of mapping service provides more flexibility and its advantages can become more significant with the increase of network scale, especially in scenarios with continuous mobility. Existing locator/identifier split architectures mainly focus on the terrestrial network. And
related work on its application feasibility in satellite networks has not been conducted \cite{Han2016b}. Table \ref{LocatorIdentComparison} compares the reviewed studies on locator/identifier split algorithms and highlights their limitations in the environment of future LEO SatNets.


\begin{table*}[t]
\caption{Comparison of locator/identifier split algorithms.} 
\centering 
\begin{tabular}{|p{1cm}|p{4cm}|p{4cm}|p{4cm}|p{4cm}|}
\hline
\cellcolor{aliceblue}\textbf{Algorithm} &  \cellcolor{aliceblue}\textbf{ Location resolver placement}	 & \cellcolor{aliceblue}\textbf{Location update frequency} & \cellcolor{aliceblue}\textbf{Location query trigger} & \cellcolor{aliceblue}\textbf{Main limitation in LEO satellites mega-constellation}\\ 
\hline 
\hline
HIP \cite{HIP2006}, \cite{HIP2017} & Ground-based& Every location change & Establishing a connection with CN or when CN change its location & Not investigated for satellite networks\\
\hline

LISP \cite{Farinacci2013} & Ground-based & When MN moves from one edge network to another & When the location is not available in ITR cache& Not investigated for satellite networks\\
\hline

GRIMM \cite{Han2016b} & Ground-based & Global update: when a MN moves from one region to another. Local update: when accessed satellite change &  When new data session starts and its done through SGLs& Global updates will create high network overhead, requires large storage to keep global records, Ground-based location resolution delays session initiation\\
\hline

SAT-GRD  \cite{Feng2016} & Ground and satellite-based & When MN moves from one AR to another within the same edge, the location update is in the edge mapping system. Whereas, moving from one edge network to another results in location updates at the core mapping system level & When moving from one AR to another or from one edge network to another & Did not consider the direct communication between satellites and users. The border routers might encounter high loads and bottlenecks when direct communication between satellites and users is considered.\\
\hline

HetNet \cite{Feng2017b} & Ground-based & Upper hierarchical level update: when MN moves from one edge network to another. Local update: when MN moves within the same edge network& When new data session starts and its done through SGLs& Did not consider the direct communication between satellites and users \\
\hline

VAP \cite{Zhang2017} & Satellite-based & When a MN or satellite changes its VAP region  &  When new data session starts and it is done through ISLs & Creates overhead on ISLs when location-ID records are transferred from one satellite to another.\\
\hline
RMRS \cite{Zhang2017b} & Satellite-based & Updates are done in the original and replica satellites when a MN or satellite changes its virtual region & When new data session starts and its done through ISLs &Creates overhead on ISLs when location-ID records are transferred from one satellite to another. Creating replicas increases the location update cost and consumes the satellite limited processing power. \\

\hline
\end{tabular}
\label{LocatorIdentComparison} 
\end{table*}

\subsection{Issues to Consider}

This section highlights the issues that should be considered in implementing the location/identifier approach for future LEO SatNets.
\begin{itemize}
    \item To enable implementing location/identifier split approach in future SatNets, it is important to have optimal location update/resolution schemes that are scalable and can work rapidly with reduced complexity and signaling costs.
    \item Placing location resolvers on the ground may create delays in location update/resolution and they might not be uniformly distributed due to the geographical structure of Earth. On the other hand, placing the location resolvers of a certain region on satellites  requires periodic transmission of location-ID mapping records from one satellite to another, which may congest the ISLs. In this regard, placing location resolvers on \acrfull{haps} may provide a good solution as the HAPS position is in between satellites and ground stations/users, and it is considered quasi-stationary with respect to Earth \cite{kurt2020} and \cite{Sahabul2020}. 
    \item In the future, satellite networks are going to be part of the integrated \acrfull{vhetnet} \cite{alzenad2019}. Therefore, any locator/identifier split system should consider the backward compatibility with legacy IP locator and identifier systems. 
    \item Global updates performed by some of the existing solutions seem impractical especially when there are thousands of satellites and millions of user devices communicating with the satellite networks.
    \item Most of the existing studies do not consider the status of future satellite networks which will consist of several mega-constellations that will provide connectivity and Internet services not only in rural or remote areas but also in urban areas.  
\end{itemize}

\section{Approach \#3: SDN-based location management in LEO SatNets} \label{Approach3}

SDN concept separates the control plane from the data plane.  In SDN, controllers are considered the brain that performs intelligent functionalities. Through the northbound interfaces, the controllers interact with applications to decide on how to create/update flow tables saved in the SDN switches. Communication among SDN controllers can be done through westbound and eastbound interfaces. Through secure channels, an SDN controller can communicate with one or more SDN switches. Actions on how to treat the received packets are predefined in the switches flow tables.  Whenever a new packet is received at the SDN switch, a flow table lookup is done. In the lookup process, if the packet headers match can be found in the lookup table, then the predefined actions are performed. If the match was not found in the flow table entries then the packet will be forwarded to the controller through the southbound interface \cite{Kaloxylos2020}. For more details on SDN, the interested reader may refer to \cite{Sood2015}.


Based on the received network topology information, the SDN controller makes the routing/forwarding decisions. Unlike traditional networks, topology management is a fundamental task in SDN. \acrfull{td} is a key component to support the logically centralized control and network management principle of SDN. TD enables a controller to have global visibility of the complete network. Discovering the network topology includes the discovery of  switches, hosts, and interconnected switches. In the TD process, each entity on the network can collect information about the network topology. The information collection can be done at different levels and in many ways. 
In addition, the TD process must be efficient in terms of sending topology information only when changes happen and not flooding the controllers with unnecessary information  \cite{Kipongo2018}. 

Based on the aforementioned explanation of SDN's TD, it can be concluded that TD is providing a major part of the functionality of location management in SDN. Basically, location management and TD both provide information on the network entities' location (logical location) within the network and the interconnections between different entities.   For this reason, several studies proposed SDN based mobility management schemes where TD is utilized for location management.  For example, \cite{Sulovic2017} proposes a location management scheme for a 5G mobile core network that purely relies on SDN. The scheme is used to manage the MN status and the paging procedure. 


In \cite{Kipongo2018}, discusses the challenges of TD protocol for SDN-based wireless sensor networks. The author highlighted the issue of limited resources of sensor networks that require lightweight TD protocol, which is a similar requirement in satellite networks. Although satellite networks will have more resources than wireless sensor networks in terms of power and processing capabilities, applying existing TD protocols in satellite networks will encounter very high network overhead. In particular, with the fast mobility of satellites that results in frequent topology changes in the densely deployed mega-constellations, more packets will be sent to the controller to update the topology and flow table. Such an
overhead traffic could negatively affect the efficiency of network resources utilization.

The authors in \cite{Zali2018} and \cite{Petropoulos2017} utilized SDN to construct and manage the topology of an \acrfull{icn} overlaid on a legacy IP network using the controller’s management capabilities. A centralized controller constructs the ICN
topology dynamically when new customer networks join the overlay, and it can modify the topology of the ICN overlay in order to reduce the load on the congested links. This dynamic topology control performed by the centralized controller is a useful feature for future LEO SatNets. However, the centralized nature of SDN will raise a concern with respect to satellite network scalability. 

In SDN, the controller is responsible for updating the forwarding rules of the network elements’ in the data plane. The time required for a rule to be installed is referred to as the flow setup time. In a large-scale network such as a LEO mega-constellation, a  single controller with limited resources will not be able to handle all the update requests originating from the data plane and the controller might encounter bottlenecks. Furthermore, due to the large distances between the satellites and the controller, there is no guarantee to meet the acceptable control plane latency. Therefore, having a distributed control plane becomes mandatory. However, it is very important to choose the optimum number of controllers and their locations based on the traffic load distribution and topology changes~\cite{Papa2018}.

\subsection{Merging of SDN and Distributed Mobility Management}

\begin{figure*}[h]
\centering
\includegraphics[width=1\textwidth]{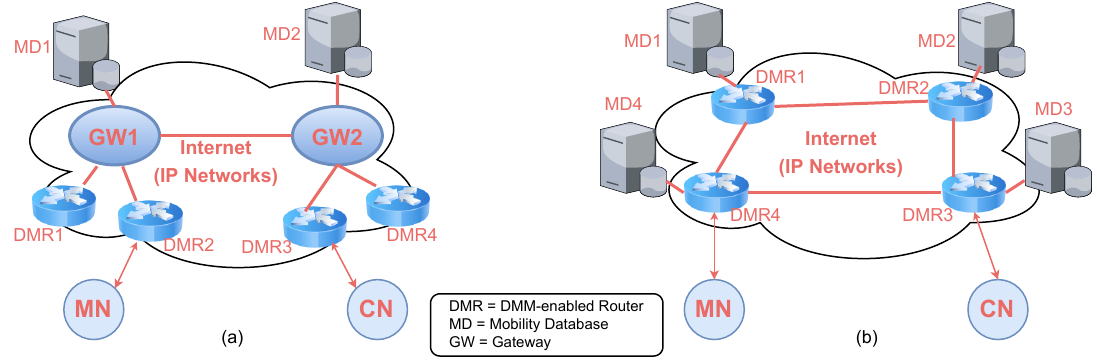}
\caption{DMM architecture. (a) Partially distributed. (b) Fully distributed.}
\label{DMMTypes}
\end{figure*}

The DMM concept was introduced by IETF to overcome the limitations of centralized management such as the one point of failure, bottlenecks, and high delays. DMM-based solutions can be categorized into two types (Figure \ref{DMMTypes}), depending on the distribution level of the control plane \cite{Cordova2019}:
\begin{itemize}
    \item \textbf{Partially distributed:} The control plane is centralized in certain control points in the network, whereas the data plane is completely distributed among the network entities; and
\item \textbf{Fully distributed:} Both control plane and data plane are completely distributed among the network entities, and there exists no central entity of control.
\end{itemize}

Some researchers considered SDN as an enabler for DMM \cite{Perras2019}. They considered that DMM approaches will push mobility anchor points to be distributed at the network edge and foreseen the SDN separation of data and control planes, which allows quicker configuration and provision of network connections, as an enabler for managing mobility in a distributed way at the edge. On the other hand, some studies considered that DMM can help in mitigating the drawbacks of SDN centralization. The merging of SDN and DDM concepts have been investigated in several types of networks. For example, an architecture that uses the SDN paradigm with the DMM concept in an environment of heterogeneous IP networks was proposed in \cite{Cordova2019}.  The proposed architecture avoids the centralization problem of SDN and presents a hierarchical cluster-based implementation of the SDN‐DMM controllers that improves the scalability of the control plane and reduces the availability problem related to the single point of failure. With the merging of SDN and DMM, a large portion of data traffic can be handled locally at the network edge, which reduces the probability of bottlenecks at the network core. When a MN moves to another network, it does not need to change its IP address while being in an ongoing session. Instead, the controller will update the IP flow related to the moving MN to ensure that the forwarded packets will reach the MN in the new network. In addition, several studies studied the merging of DMM and SDN in 5G \cite{Hakimi2018}, \cite{Nguyen2016}.

\subsection{SDN-based Location Management in LEO SatNets }
A simple \acrfull{sdsn} architecture was proposed in \cite{Bao2014}. It contains three planes: the data plane (satellite infrastructure, terminal router), the control plane (a group of GEO satellites), and the management plane (\acrfull{nocc}). Similarly, the author of \cite{Li2017} proposed a SDSN where the controllers are located on GEO satellites and the switches are deployed in MEO and LEO satellites. Since the frequent handovers will rapidly increase the flow table size in SDSN,  a lot of flows will be dropped during topology changes (handovers) due to the limited size of the flow table. In particular, a commodity switch can store about 1500 entries only  because of the high cost and energy consumption of the \acrfull{tcam} that is usually used by SDN switches \cite{Curtis2011}. After a handover, the flow table will have unexpired entries occupy the flow table space and are useless because they will not be used any more. The subsequent flows may be dropped if the TCAM space is limited. To address this problem the author of \cite{Li2017} proposed a heuristic \acrfull{tsmm} algorithm which aims to reduce the drop-flow during handover. The TSMM algorithm adjusts the entries timeout dynamically while considering two key points, the limited flow table space and the satellite link handover. This aims to discard the flow entries that belong to the former connection when the handover occurs. This work has been extended in SAT-FLOW \cite{Li2017b}, which is a multi-strategy flow table management method for SDSN. SAT-FLOW is composed of two heuristic algorithms, \acrfull{dct} algorithm and TSMM algorithm. DCT calculates a dynamic idle timeout value for the flow entries taking limited TCAM space and classified traffic into consideration. TSMM utilizes the result of DCT and considers link handover in satellite networks. Thus, DCT aims to reduce the flow table size and TSMM aims to reduce the drop-flows during the handover. A time estimation model is proposed by \cite{Boero2018}  to estimate the mean time required to complete the SDN control (i.e., finding or creating the necessary traffic flow) and to deliver the first packet to the destination. The author considered an architecture where three GEO satellites play the role of SDN controllers and the data plane is distributed among several LEO satellites. 

However, the fixed controller placement at GEO satellites might not be able to react to traffic fluctuations caused by variable user activities and different time zones. In addition, with densely populated and highly dynamic LEO mega-constellations, having few controllers placed at GEO may result in bottlenecks and high delays while updating routing flows.

To overcome the fixed placement problem, a dynamic SDN controller placement is considered in \cite{Papa2018}. The author developed a mathematical model and formulated it as an \acrfull{ilp} to find the optimal controller placement and the number of satellites that will work as controllers. The goal of the model is to minimize the average flow setup time with respect to the traffic dynamics. The model was derived based on an SDN architecture, where the control plane layer consists of several LEO satellites that varies based on traffic demands in addition to seven satellite gateways placed on the ground and serve as entry points to the backbone network. The gateways position is fixed and is obtained from the existing IRIDIUM system. The satellites that are part of the control plane serve as both controllers and network switches. They manage, control, and update the forwarding rules of the flow tables of the satellites of the data plane. On the other hand, the satellites of the data plane are only responsible for forwarding packets based on rules defined by the corresponding controllers. However, this study considered the LEO satellites in the number of hundreds, which does not reflect the mega-constellation characteristics of future networks. The future densely deployed satellite networks may result in creating huge flow tables that have very short lifetime.

A framework of SDSN is proposed by \cite{Shuai2018}, which defines the \acrfull{dcpp} as well as the \acrfull{scpp}, and used an \acrfull{apso} algorithm to solve the problems in DCPP and SCPP. The author considered the parameters of propagation delay, reliability, controller load, signalling cost, and the dynamic characteristic of satellite networks. In the DCPP the number and location of active controllers can be adjusted based on the changes in network conditions. The author assumed an architecture of multiple  controllers that can  be placed in LEO, MEO, or GEO, whereas the switches are all placed in the same LEO constellation. In this work, the data plane switches send periodic hello messages to their controller, such messages are used to collect information about changes in network topology (i.e., topology discovery). In addition, the periodic hello messages are used to connect with a new controller when the switch loses the connection with the current controller. Controllers exchange the obtained topology discovery information to build up a global view of the network.

A three-layer hierarchical controller architecture for software-defined GEO/LEO satellite networks is proposed in \cite{Xu2018b}. The solution exploited the wide coverage ability of GEO satellites, easy upgrade and maintenance of NOCC, and stability of inter-satellite links in the same low earth orbit. The control plane consists of domain controllers, slave controllers, and a super controller. The GEO satellites are set as domain controllers because of their broadcast capabilities over a wide-coverage area and stable connection with the ground station. The domain controller monitors and manages the LEO satellites in its coverage. The LEO satellites forward and collect the network status information, and are divided into different domains according to the GEO coverage. Several slave controllers are selected from LEO satellites. The GEO domain controllers communicate with just the slave controllers under their own authority instead of all LEO satellites in their domain. By using inter-satellite links, the slave controllers collect the status information of the LEO satellites under their own authority, which is then sent to the corresponding domain controllers. 
The NOCC is deployed as a super controller that can obtain the knowledge of the overall network through the primary GEO satellites. Based on the aforementioned description, a logically centralized control plane with global knowledge is created through physically distributed LEO controllers. To select the slave controller, each LEO orbit plane is regarded as a separate management area. The LEO satellites in each orbit plane are divided into two groups such that one group is moving from north to south and the other from south to north. 
In each group, the LEO satellite whose latitude is nearest to $0^{\circ}$ is selected as the slave controller to collect the status information of other LEO satellites in its group. However, if more than one LEO satellite has the same lowest latitude, then the slave controller is chosen as the one which can maintain a longer communication time window.
Table \ref{TableSDN} compares the aforementioned studies and highlight their limitations in the context of future LEO SatNets.




\begin{table*}[t]
\caption{Comparison of SDN based location management in LEO SatNets.} 
\centering 
\begin{tabular}{|p{1.4cm}|p{3cm}|p{3cm}|p{2cm}|p{3cm}|p{3cm}|}
\hline
\cellcolor{aliceblue}\textbf{Algorithm} &  
\cellcolor{aliceblue}\textbf{Study objective}	 &
\cellcolor{aliceblue}\textbf{Controller placement}	 &
\cellcolor{aliceblue}\textbf{Dynamic/Fixed} & 
\cellcolor{aliceblue}\textbf{Location update frequency} &
\cellcolor{aliceblue}\textbf{Main limitation in LEO satellites mega-constellation}\\ 
\hline 
\hline
OpenSAN \cite{Bao2014} & Proposed a SDN-based satellite network architecture & In GEO satellites & Fixed & Based on satellite movement prediction, flow tables are updated & This architecture is designed to provide services for terrestrial users through ground gateways only and not through direct communication. \\
\hline
TSMM \cite{Li2017}, DCT \cite{Li2017b} & To manage flow tables and reduce the drop-flows due to frequent handover & In GEO satellites & Fixed & With every handover & The authors presented a good idea to manage flows tables, however, the focus was on satellites only without considering terrestrial terminals or users.\\
\hline
Time estimation model\cite{Boero2018} & To estimate the mean time required to complete the SDN control and to deliver the first packet to destination & In GEO satellites & Fixed & Periodically based on satellite movement & Communicate with terrestrial networks using ground gateways. No consideration for the case when thousands of users are directly connected to LEO satellites.\\
\hline
Dynamic controller placement-ILP \cite{Papa2018} & To present a mathematical model that finds the optimal controller placement and the number of satellites that will work as controllers & In LEO satellites (with variable number and placement) & Dynamic & Based on the satellite movement prediction & Due to LEO satellite fast movement, SDN controllers will change frequently. It is very complicated to manage the changes in controllers seamlessly in a network consisting of thousands of satellites while serving millions of users.\\
\hline
Dynamic controllers placement- Swarm optimization \cite{Shuai2018} & To present an architecture of multiple  controllers placed dynamically using swarm optimization & In GEO, MEO, LEO satellites & Dynamic & Using periodic hello messages & Using periodic hello messages will consume the ISLs resources especially in large scale networks.\\
\hline
Hierarchical dynamic controller placement \cite{Xu2018b} & To present a hierarchical dynamic controller architecture & In GEO and LEO satellites & Dynamic & Periodically collected and sent by LEO controllers & Maintaining the hierarchical architecture in a very dense satellite network consumes ISLs resources due to the very frequent topology changes. \\
\hline
\end{tabular}
\label{TableSDN}
\end{table*}

\subsection{Issues to Consider}

This section points out the important issues that should be considered in order to utilize SDN-based location management for future LEO SatNets.

\begin{itemize}
    \item In software defined future LEO SatNets, there will be millions or billions of user devices connected to LEO satellites. This will create a huge number of flow records at each switch (LEO satellite), and such records will be expired once the satellite moves to serve a different group of people. Setting up new flow records with every satellite handover consumes resources and creates delays. However, the idea of relaying flow tables from a departing satellite to a coming one worth investigation. 
    \item When terrestrial and satellite networks are integrated, there will be millions or billions of flows. Storing, maintaining, and searching through these flow tables records is a complicated and critical issue. 
    \item Although the distributed SDN architecture is preferred in SatNets environment, the limited satellite resources to implement all the distributed SDN functions should be taken into consideration.
    \item To gain the advantages of dynamic SDN controller placement in future LEO SatNets, a number of factors should be considered such as traffic demands, users distribution, users mobility, signalling cost, and the dynamic characteristic of satellite networks.
    \item Future network management automation merges the concept of SDN with artificial intelligence/machine learning \cite{Tasneem2020}. Thus, the utilization of artificial intelligence in SDN-based LEO SatNets location management should be considered. In particular, artificial intelligence/machine learning algorithms can be useful in selecting the SDN controllers and their placements in order to adapt to the dynamic nature of LEO SatNets.
    \item In terrestrial networks, SDN controllers update the flow tables using the information obtained through topology discovery protocols. In satellite networks, a large portion of topology update overhead can be saved by predicting satellite movement. However, with the availability of mega-constellations of thousands of satellites, user devices will have multiple candidate satellites to handover. In this situation, updating users' related flows based on mobility predictions is complicated.
\end{itemize}

\section{\textcolor{black}{Current view and future Directions}}\label{Currentview}

The previous three sections focused on discussing the efforts that have been done to address the problem of location management in LEO SatNets. Some researchers tackled the issue of LEO SatNets location management  by extending or enhancing the IETF IP-based location management techniques that come as part of IPv6 mobility management protocols. In contrast, a considerable number of studies identified the dual role of IP address as an identifier and a locator to be the main cause of the poor performance of IETF IP-based location management techniques in LEO SatNets. Consequently, several researchers investigated the employment of the location/identifier split concept in LEO SatNets. A third approach utilized the network softwarization and the decoupling of control and data planes advantages offered by SDN to support location management in LEO SatNets in a flexible way.

Clearly, the IETF IP-based location management techniques and location/identifier split algorithms are based on two totally different concepts as the former considers the dual role of IPv6 address while the later separates the two roles. IPv6 and SDN are interrelated technologies where IPv6 operates on the network layer and SDN handles the management of the networking operations. From the location management perspective, SDN serves more the routing side by using the flow concept rather than IP addresses to deliver packets to their destinations. 

Under each of the three aforementioned approaches, several solutions have been proposed to deal with location management in LEO SatNets. Although such solutions have some potentials when applied to future LEO SatNets, many challenges will be encountered as well. This is due to the complicated and new mobility and topology characteristics of future LEO SatNets, as discussed in Section \ref{CharaAndChallenge}.

\begin{table*}[t]
\caption{Summary of three approaches' advantages and challenges when applied to future LEO SatNets.} \label{CompariosonTable}
\centering 
\begin{tabular}{|p{3cm}|p{5cm}|p{6cm}|}
\hline
\cellcolor{aliceblue}\textbf{Approach} &  
\cellcolor{aliceblue}\textbf{Advantages}	 &
\cellcolor{aliceblue}\textbf{Challenges}\\ 
\hline 
\hline
Approach \#1: Extensions of IETF location management techniques & 
\begin{itemize}
    \item Easy to deploy as they are extensions of the IETF IPv6 standardized mobility management protocols.
    \item Easy to ensure compatibility with other IP-based networks (even with IPv4-based networks).
\end{itemize}
  & \begin{itemize}
    \item Since the IP address is used as a locator as well, frequent IP address changes will occur in the highly dynamic environment of future SatNets where thousands of users, mobile BS (satellite), routers are changing their access point every moment. This will result in high signaling cost, delays, and packet loss.
    \item Depending on IP addresses for data packets forwarding will result in non-optimized routing, which will extremely degrade the network performance especially when ground to space communication links resources are used for dispensable routes.
    \item The performance of the proposed extensions will be severely affected by the placement of the nodes that will play the anchor role in the location management of future LEO SatNets.
\end{itemize}  
\\
\hline

Approach \#2: Locator/identifier split & 
\begin{itemize}
    \item Terminals and network entities can keep their identifiers and they need to update their location as they change their point of access. 
    \item As data packets are forwarded to the destination logical location rather than its IP address, more optimized routing can be achieved. 
\end{itemize}  & 
\begin{itemize}
    \item This approach might face some incompatibility issues with IP-based networks.
    \item It requires an efficient location update/resolution system that can handle a rapidly changing topology where millions of users and network devices are involved. The system should be scalable and provide fast responses with low complexity and signaling costs.
    \item The placement and the architecture of the location resolution system is a very challenging issue in future LEO SatNets, where distance and restrictions on link budget affect communication. 
\end{itemize}
\\
\hline

Approach \#3: SDN-based location management &
\begin{itemize}
    \item SDN concept adds the programmability feature to network management which supports agility and flexibility.
    \item SDN supports policy-driven network management and network automation.
\end{itemize} & 
\begin{itemize}
    \item It is foreseen that the merging of SDN and DMM concepts will support the scalability of future LEO SatNets. However, this will come with the price of increasing the complexity of the control and management planes due to following the distributed architecture instead of centralized. 
    \item In software defined future LEO SatNets, controllers placement is a critical issue. 
    \item With the large number of user devices and network entities, flow tables storage and maintenance require new techniques that can handle the frequent topology changes that happen in large volumes. 
\end{itemize} 
\\
\hline

\end{tabular}
\end{table*}

\textcolor{black}{For IP-based solutions, the main challenge is to minimize and manage the consequences of frequent IP address change (e.g., signaling cost, delays, and packet loss), non-optimized routing, and inefficient placement of physical or logical anchors (i.e., a home agent). With respect to the second approach, the locator/identifier split, the main challenges are the back compatibility with existing IP-based networks, the scalability of location update/resolution systems, and the placement of the location resolution system. Although the SDN-based location management approach is promising for future LEO SatNets, it will encounter some challenges. Managing a distributed SDN control plane over a large-scale network is complex due to the controller selection/placement and the critical processes related to managing the rapidly changing flow tables in the highly dynamic environment of future LEO SatNet.} Table \ref{CompariosonTable} summarizes the advantages and challenges of each of the three location management approaches from the perspective of future LEO SatNets.




Intensive research is required to overcome the obstacles and unlock the potentials of future LEO SatNets to providing continuous connectivity everywhere, for everything, and in the required QoS. 
The following are some critical points that require further investigation to realize future SatNets. 
\begin{itemize}

    \item In future LEO SatNets, there will be several mega-constellations with different orbital parameters. Such parameters will  have an effect on a number of variables including propagation delays, handover frequency and duration, footprints, and density of satellites. Such variables affect the performance of location management algorithms. Thus, for different orbits/constellations, location management  might be different. In addition, the diversity in required QoS for user devices/applications should be taken into consideration while designing location management solutions for future LEO SatNets. 
    
    \item \textcolor{black}{Advances in communication technologies will enable direct communication between satellites and small devices with limited power (e.g., mobile phones and sensors). We envision a frequent utilization of future LEO SatNets communication services in highly populated areas. Providing services to rural and remote areas may not be economically viable by itself and satellite networks operators will likely elect to also provide services in urban areas with high user density to improve market penetration and close their business case. However, most of the existing research on location management in LEO SatNets focuses on the cases with a low density of users (e.g., users in rural or remote areas) or indirect communication with satellites through ground gateways.   Therefore, to support future LEO SatNets, location management schemes should be designed to handle thousands or millions of devices connected directly to satellites. In such a scenario, issues of address resolution/ mapping, flow tables management, and handovers of a large group of users should be considered and investigated.}
    
    \item \textcolor{black}{Future networks are expected to be self-evolving networks (SENs) that utilize artificial intelligence to make future integrated networks fully automated and intelligently evolve with respect to the provision, adaptation, optimization, and management aspects of networking, communications, computation, and infrastructure nodes’ mobility \cite{Tasneem2020}. To work in the self-evolving environment of future LEO SatNets, location management systems should be able to self-restructure the network logical topology to improve network performance. In this regard, the technologies of SDN and network slicing have promising potential. However, this topic requires further research.}
    
    \item The authors in \cite{Zhang2016} considered that the main causes of the current Internet’s problems are the so-called triple bindings, namely user/network binding, control/data binding, and resource/location binding. The author proposed a collaborative Internet architecture that completely cancels the restrictions imposed by the triple bindings. Although this approach applicability in future LEO SatNets was not discussed, it worths the investigation as it may add  flexibility to network topology management.  
    
    \item As part of the secure mobility management work presented in \cite{Lai2019}, the author proposes to use blockchain technology for group location management in vehicular ad hoc networks. Blockchain technology is well known for managing its ledgers in a secure and distributed nature. This feature of blockchain might be advantageous in managing the flow tables in SDN-based LEO SatNets.
    
    \item Recently, discussions have started on proposing a \textit{``New IP Address"} for 2030 networks \cite{Chen2020}, \cite{Jiang2020b}, \cite{Kheirkhah2020}, which aims to connect heterogeneous networks, provide deterministic forwarding, and support intrinsic security. Theoretically, New IP offers more efficient addressing and network management than the existing TCP/IP standard. However, there are some concerns that New IP would require authorization and authentication of the sent data packets and the user identity as well as the internet addresses.
    

\end{itemize}






\section{Conclusion}\label{con}

Efficient location management is essential to unlocking the potential of future LEO SatNets. This article aims to explore the current development in location management and identify the gaps and challenges facing the realization of required location management for future LEO SatNets. From the perspective of future LEO SatNets, this article critically reviews the existing three location management approaches including extensions of the IETF location management techniques approach, the locator/identifier split approach, and the SDN-based location management approach. The deterministic aspects of the LEO mega-constellations and the fixed terminals can be exploited to support location management. Software defined satellite network concept will play a major role in supporting location management in future LEO SatNets. Worthy recommendations for future research directions conclude this work. This article can be used as a road map to guide the research efforts towards fulfilling the requirements of location management in future LEO SatNets environment.

\bibliographystyle{IEEEtran}
\bibliography{ref}

\begin{IEEEbiography}[{\includegraphics[width=1in,height=1.25in,clip,keepaspectratio]{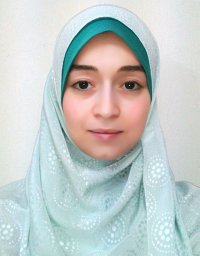}}]{Tasneem Darwish} is currently a postdoctoral fellow at the Department of Systems and Computer Engineering, Carleton University, Canada.  She received the MSc. degree with merit in Electronics and Electrical Engineering from the University of Glasgow, UK, in 2007 and her Ph.D. degree in Computer Science from Universiti Teknologi Malaysia (UTM), Malaysia, in 2017. From  2017 to 2019, Tasneem was a postdoctoral fellow at UTM . From 2019 to 2020, she was a research associate at Carleton University, Canada. In 2020, Tasneem started working as a postdoctoral fellow at Carleton University on a collaborative project with MDA Space to investigate mobility management in future LEO satellite networks. She is the recipient of the UTM Alumni Award for Science and Engineering in 2017. She was awarded the Malaysia International Scholarship (MIS) from 2013 to 2016. Her current research interests include mobility management in future LEO satellite networks, edge/fog computing and data offloading in HAPS, vehicular ad hoc networks, and intelligent transportation systems.  Tasneem is a senior IEEE member and an active reviewer for several IEEE journals such as IEEE Internet of Things, IEEE Access, IEEE Transactions on Vehicular Technology, and IEEE Transactions on Intelligent Transportation Systems. 
\end{IEEEbiography}

\begin{IEEEbiography}[{\includegraphics[width=1in,height=1.3in]{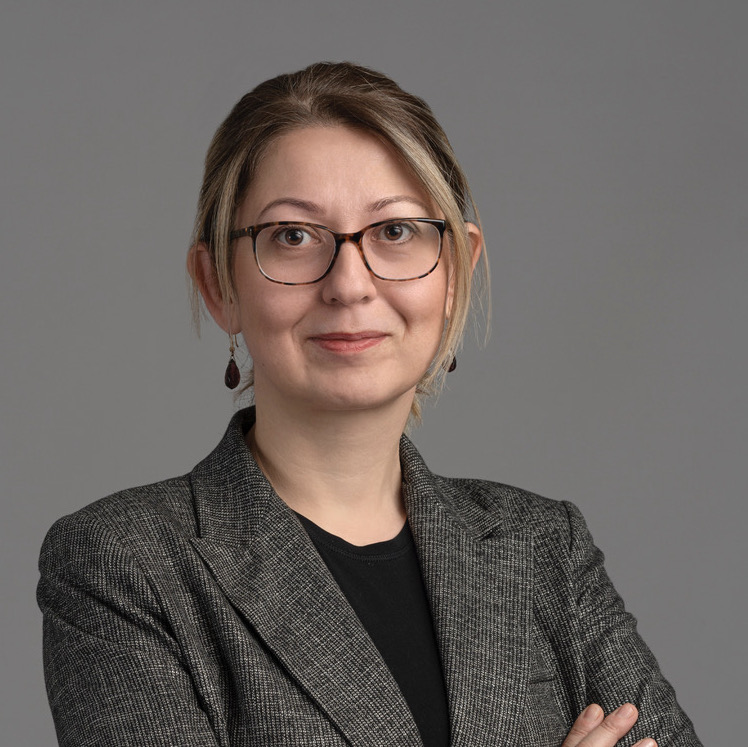}}]{Gunes Karabulut Kurt} is currently an Associate Professor of Electrical Engineering at Polytechnique Montréal, Montreal, QC, Canada.  She received the B.S. degree with high honors in electronics and electrical engineering from the Bogazici University, Istanbul, Turkey, in 2000 and the M.A.Sc. and the Ph.D. degrees in electrical engineering from the University of Ottawa, ON, Canada, in 2002 and 2006, respectively. From 2000 to 2005, she was a Research Assistant at the University of Ottawa. Between 2005 and 2006, Gunes was with TenXc Wireless, Canada. From 2006 to 2008, she was with Edgewater Computer Systems Inc., Canada. From 2008 to 2010, she was with Turkcell Research and Development Applied Research and Technology, Istanbul. Gunes has been with Istanbul Technical University since 2010, where she is currently on a leave of absence. She is a Marie Curie Fellow and has received the Turkish Academy of Sciences Outstanding Young Scientist (TÜBA-GEBIP) Award in 2019. She is an Adjunct Research Professor at Carleton University. She is also currently serving as an Associate Technical Editor (ATE) of the IEEE Communications Magazine and a member of the IEEE WCNC Steering Board.
\end{IEEEbiography}

\begin{IEEEbiography}[{\includegraphics[width=1in,height=1.3in]{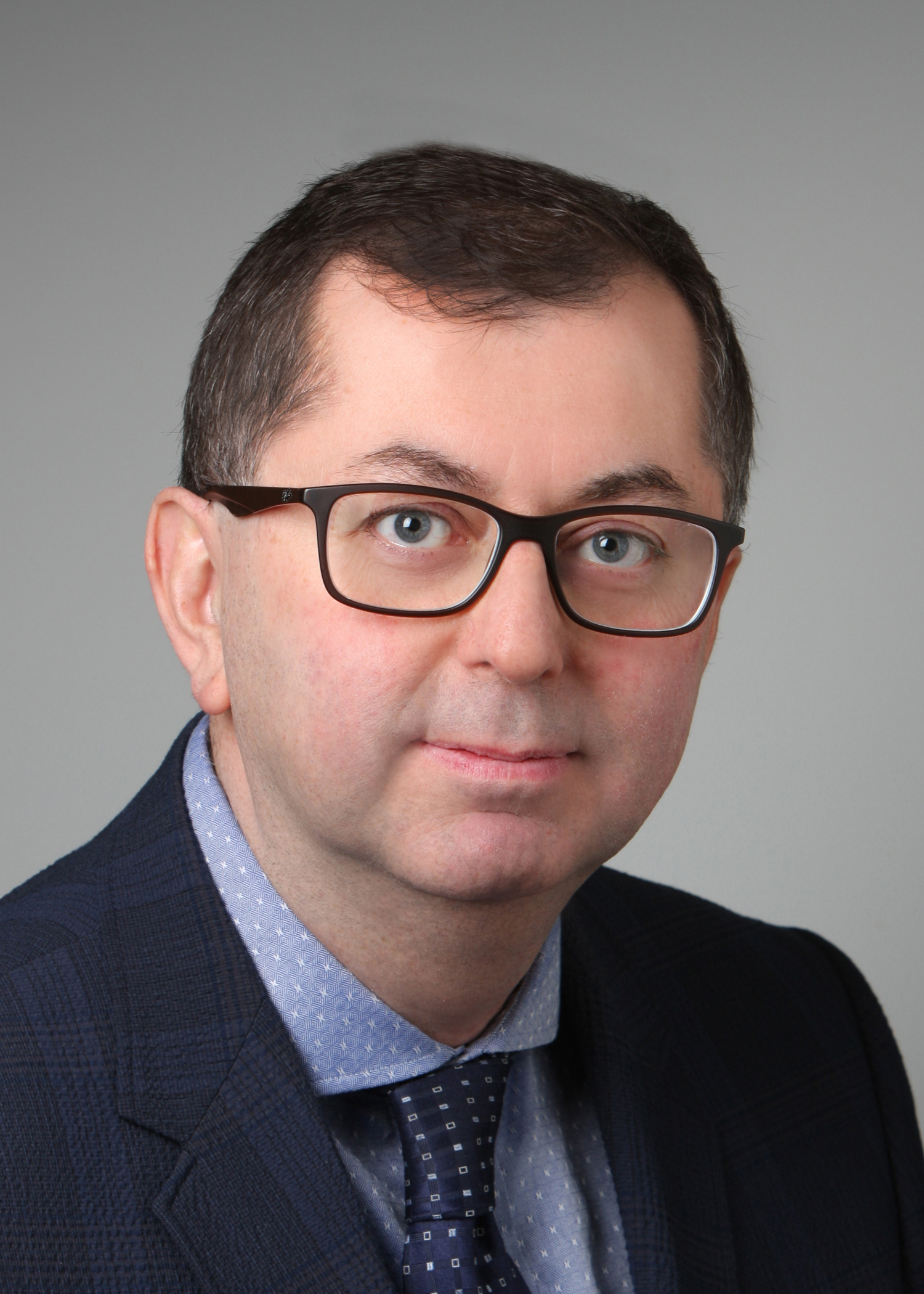}}]{Dr. Halim Yanikomeroglu} is a Professor in the Department of Systems and Computer Engineering at Carleton University, Ottawa, Canada. His primary research domain is wireless communications and networks. His research group has made substantial contributions to 4G and 5G wireless technologies. His group’s current focus is the aerial (UAV and HAPS) and satellite networks for the 6G and beyond-6G era. His extensive collaboration with industry resulted in 37 granted patents. He is a Fellow of IEEE, EIC (Engineering Institute of Canada), and CAE (Canadian Academy of Engineering), and a Distinguished Speaker for both IEEE Communications Society and IEEE Vehicular Technology Society. He is currently serving as the Chair of the IEEE WCNC (Wireless Communications and Networking Conference) Steering Committee. He served as the General Chair or TP Chair of several conferences including three WCNCs and two VTCs. He also served as the Chair of the IEEE’s Technical Committee on Personal Communications. Dr. Yanikomeroglu received several awards for his research, teaching, and service, including the IEEE ComSoc Fred W. Ellersick Prize in 2021, IEEE VTS Stuart Meyer Memorial Award in 2020, and IEEE ComSoc Wireless Communications Technical Committee Recognition Award in 2018.
\end{IEEEbiography}

\begin{IEEEbiography}[{\includegraphics[width=1in,height=1.3in]{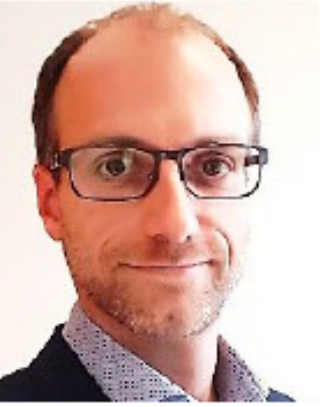}}]{Guillaume Lamontagne} received the B.Eng. and M.Eng. degrees from the École de Technologie Supérieure (ÉTS), Montréal, QC, Canada, in 2007 and 2009 respectively. His experience in satellite communications started through internships and research activities with the Canadian Space Agency (CSA), in 2005, and the Centre national d’études spatiales (Cnes), France, in 2006 and 2008. He joined MDA in 2009 and held various communication systems engineering and management positions before being appointed as the Director of Technology, Payloads, in 2019. Through this role, he is leading MDA’s Research and Development activities for satellite communications as well as establishing the related long term development strategy.
\end{IEEEbiography}

\begin{IEEEbiography}[{\includegraphics[width=1in,height=1.3in]{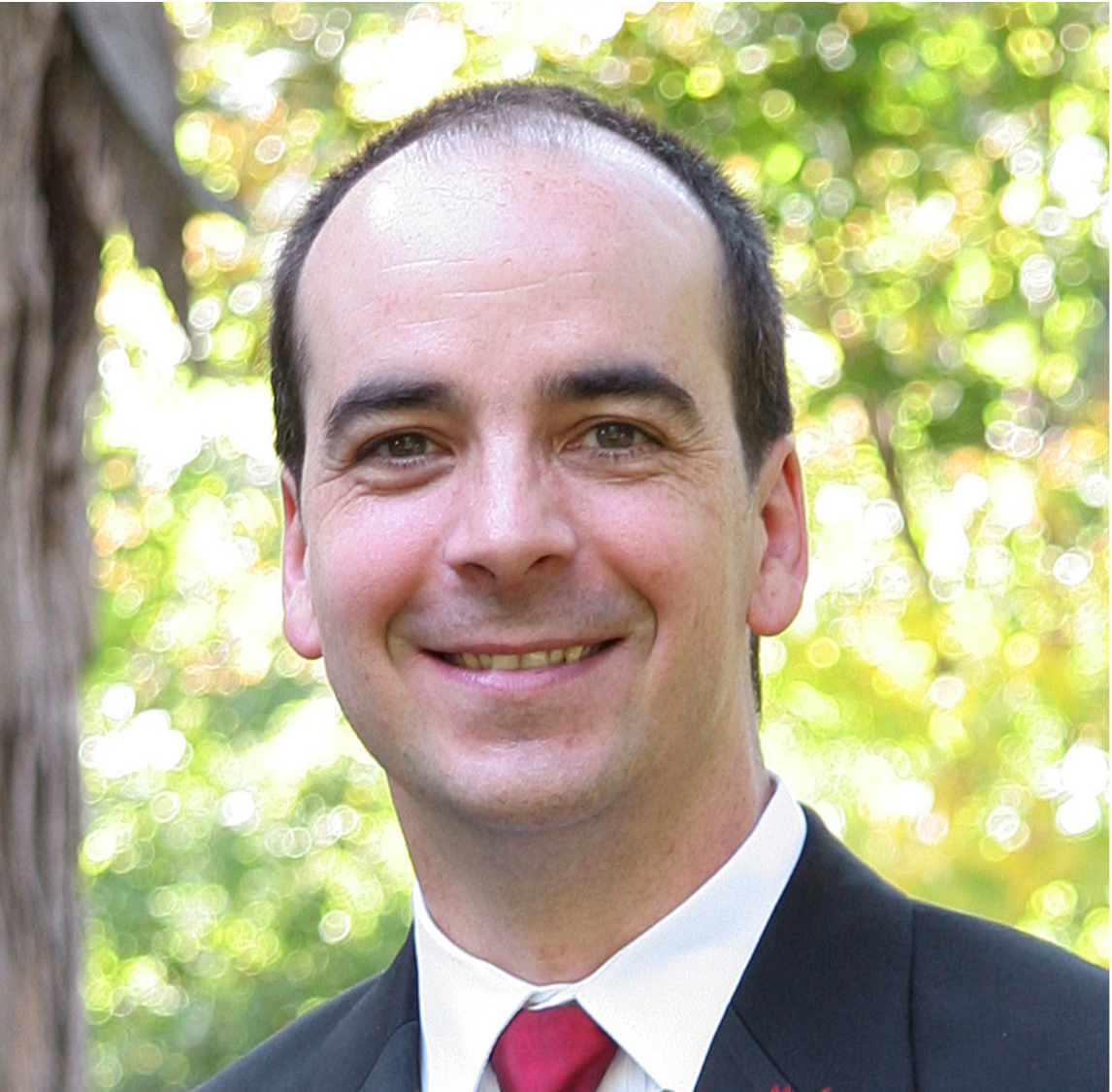}}]{Michel Bellemare}received is B.Eng. degree from Université de Sherbrooke, Sherbrooke, Québec, Canada in Communications Engineering in 1986. He has gained experience in terrestrial and space wireless communications in various companies such as Nortel Networks, Ultra Electronics, SR Telecom, etc. He is now a Space Systems Architect at MDA corporation, Sainte-Anne-de-Bellevue, Québec, Canada.
\end{IEEEbiography}

\end{document}